\documentclass[11pt]{article}
\usepackage[margin=1.1in]{geometry}              
\usepackage{graphicx}
\usepackage{amssymb}
\usepackage{amsmath}
\usepackage{epstopdf}
\usepackage{color}
\usepackage{xcolor}
\usepackage{txfonts}

\usepackage{multicol}
\usepackage{caption}
\usepackage{soul}
\usepackage{xcolor}
\usepackage{lipsum}

\DeclareGraphicsExtensions{.eps,.pdf,.jpeg,.png}

\newcommand{\w}{w}
\newcommand{\Iin}{I_{NaP}^{IN}}
\newcommand{\Imn}{I_{NaP}^{MN}}
\newcommand{\Iappin}{I_{ton}}%{I_{app}^{IN}}
\newcommand{\sighin}{\sigma_h^{IN}}
\newcommand{\sighmn}{\sigma_h^{MN}}
\newcommand{\thethin}{\theta_h^{IN}}
\newcommand{\thethmn}{\theta_h^{MN}}
\newcommand{\hinfin}{h_{\infty}^{IN}}
\newcommand{\hinfmn}{h_{\infty}^{MN}}
\newcommand{\gappp}{g_{ton,Pro}}
\newcommand{\gappr}{g_{ton,Ret}}
\newcommand{\gappl}{g_{ton,Lev}}
\newcommand{\gappd}{g_{ton,Dep}}
\newcommand{\gappe}{g_{ton,Ext}}
\newcommand{\gappf}{g_{ton,Flx}}
\newcommand{\Isynin}{I_{syn}^{IN}}
\newcommand{\Isynmn}{I_{syn}^{MN}}
\newcommand{\gsynmn}{g_{syn}^{MN}}
\newcommand{\gsynin}{{g_{j\multimap i}}} 
\newcommand{\gsyninij}{{g_{i\multimap j}}} 
\newcommand{\gsynRP}{{g_{R\multimap P}}}
\newcommand{\gsynPR}{{g_{P\multimap R}}}
\newcommand{\gsynDL}{{g_{D\multimap L}}}
\newcommand{\gsynLD}{{g_{L\multimap D}}}
\newcommand{\gsynFE}{{g_{F\multimap E}}}
\newcommand{\gsynEF}{{g_{E\multimap F}}}
\newcommand{\gsynDP}{{g_{D\multimapdot P}}}
\newcommand{\Isynexc}{{I_{exc}}}
\newcommand{\gexji}{{g_{j\rightarrow i}}}
\newcommand{\gexLP}{{g_{L\rightarrow P}}}
\newcommand{\gexDR}{{g_{D\rightarrow R}}}
\newcommand{\gexLE}{{g_{L\rightarrow E}}}
\newcommand{\gexDF}{{g_{D\rightarrow F}}}
\newcommand{\gexED}{{g_{E\rightarrow D}}}
\newcommand{\gexFL}{{g_{F\rightarrow L}}}
\newcommand{\gexFF}{{g_{F\rightarrow F}}}
\newcommand{\gexFE}{{g_{F\rightarrow E}}}

\newcommand{\Isyninh}{{I_{inh}}}
\newcommand{\ginji}{{g_{j\multimapdot i}}}
\newcommand{\ginLR}{{g_{L\multimapdot R}}}
\newcommand{\ginLF}{{g_{L\multimapdot F}}}
\newcommand{\ginDP}{{g_{D\multimapdot P}}}
\newcommand{\ginDE}{{g_{D\multimapdot E}}}
\newcommand{\ginFE}{{g_{F\multimapdot E}}}
\newcommand{\ginFF}{{g_{F\multimapdot F}}}
\newcommand{\ginFD}{{g_{F\multimapdot D}}}
\newcommand{\ginEL}{{g_{E\multimapdot L}}}

\newcommand{\sexc}{{g_{exc}}}
\newcommand{\sinhb}{{g_{inh}}}

\newcommand{\Einh}{{E_{inh}}}
\newcommand{\Eexc}{{E_{exc}}}

\graphicspath{{./Figures/}}

\title{Rhythm Generation, Robustness,  
and Control\\ in Stick Insect Locomotion: Modeling and Analysis}
\author{Zahra Aminzare 
\thanks{Department of  Mathematics, University of Iowa,  Iowa City, IA 52242 USA,
zahra-aminzare@uiowa.edu}
\; and \;   Jonathan E. Rubin
\thanks{
Department of Mathematics, University of Pittsburgh, Pittsburgh, PA 15260 USA,
jonrubin@pitt.edu}
}

\date{}                                          

\begin{document}
\maketitle

\begin{abstract}
Stick insect stepping patterns have been 
studied for insights about locomotor rhythm generation and control, because the underlying neural system is relatively accessible experimentally and produces a variety of rhythmic outputs.  Harnessing the experimental identification of effective interactions among neuronal units involved in stick insect stepping pattern generation, previous studies proposed computational models 
simulating aspects of stick insect locomotor activity. While 
these models generate diverse stepping patterns and transitions between them, 
there has not been an in-depth analysis of the mechanisms underlying their dynamics.
In this study, we focus  on modeling rhythm generation  by the neurons associated with the protraction-retraction, levitation-depression, and extension-flexion antagonistic muscle pairs of the mesothoracic (middle) leg of stick insects. Our model features a reduced central pattern generator (CPG) circuit for each joint and includes synaptic interactions among the CPGs; we also consider extensions such as the inclusion of motoneuron pools controlled by the CPG components. 
The resulting network is described by an 18-dimensional system of ordinary differential equations.  We use fast-slow decomposition, projection into interacting phase planes, and a heavy reliance on input-dependent nullclines to analyze this model.  Specifically, we identify and elucidate dynamic mechanisms capable of generating a stepping rhythm, with a sequence of biologically constrained phase relationships, in a three-joint stick insect limb model.  Furthermore, we explain the robustness to parameter changes and tunability of these patterns.  
In particular, the model allows us to identify possible mechanisms by which neuromodulatory and top-down effects could tune stepping pattern output frequency.
\end{abstract}

\textbf{Keywords} central pattern generator, reciprocal inhibition,   multiple timescale dynamics, escape, release, robustness

\bigskip

%~~~~~~~~~~~~~~~~~~~~~~~~~~~~~~~~
\section{Introduction}\label{sec:introduction}

Circuits of neurons called central pattern generators (CPGs) are thought to underlie a variety of repetitive behaviors involving multiple alternating phases of activity, such as respiration (with inspiratory, post-inspiratory, and expiratory phases), multi-phase digestive rhythms, locomotor patterns, and possibly even more complex movements \cite{grillner1985nbr,marder2000mpg,marder2001cpg,grillner2006bpg,Ijspeert_CPG_review,smith2009,feldman2013,Bucher2015,berkowitz2019,steuer2019,mantziaris2020,Nirody2023}. 
CPG circuits located in the brain stem, in the spinal cord in vertebrate animals and in the ventral nerve cord in invertebrates, and in other deep brain structures serve as hubs for rhythm generation and control, and their outputs can drive the motoneurons that activate muscles and can provide coordination signals to other interneurons.  
Because of the difficulty in accessing these structures for experimental recordings, identifying specific neuron groups and interactions involved in locomotor movements has remained a challenge.  
Progress has been made, however, especially in species such as various types of insects that feature relatively less complex and accessible nervous systems while also generating an interesting range of joint and limb coordination
patterns associated with locomotor behaviors \cite{buschges2008,bidaye2018,mantziaris2020}.  
Among this group, the stick insect, {\em Carausius morosus}, has been the focus of much experimental investigation, which has been complemented by a range of computational modeling efforts investigating aspects of both intralimb and interlimb interactions in locomotor pattern generation and control \cite{ekeberg2004,daun2011,dauntoth,codianni2020,strohmer2022}. 

Stick insect legs comprise five or more segments, each with a
corresponding joint and antagonistic muscle pair \cite{bidaye2018}.
Here, we will focus on the three primary joints of the insect leg (Figure~\ref{fig:CIRCUIT}(left panel)), which are the thorax-coxa joint, responsible for horizontal protraction and retraction, the coxa-trochanter joint, which produces vertical levation and depression, and the femur-tibia joint, which controls extension and flexion of the tibia.
Experimental evidence suggests that each of these joints 
relies on a dedicated CPG unit, in line with the unit pattern generator concept \cite{grillner2006}, featuring effectively inhibitory interactions that help to maintain alternating muscle activation during behavior \cite{buschges1998,akay2004,ludwar2005,buschges2008,bidaye2018,mantziaris2020}; we will refer to the neurons associated with a single such unit as a {\em joint block}.
Over a sequence of papers, B\"{u}schges and collaborators compiled a detailed influence diagram illustrating the effective interactions between the neurons -- both within the same joint block and across joint blocks -- associated with the multiple segments within each stick insect limb (reviewed in \cite{bidaye2018}).  These interactions include those relating to direct synaptic connections as well as indirect effects related to sensory organ signals associated with various mechanical events triggered by activation of neural pools.  
The fundamental features of this influence diagram consist of: (1) inhibitory interactions between neurons within each joint block associated with the antagonistic muscle pairs for each segment, (2) patterns of on-off activation associated with each of four phases of a locomotor cycle, representing swing and stance postures, each subdivided into two parts, and (3) inferred pathways representing the effects that activation of the neurons linked with each muscle have on the activation of other such neuronal muscle groups. 
Although this diagram is highly informative, many important properties of the interactions that it specifies, including their intensities and specific impacts on target dynamics, remain undetermined.
Thus, a natural approach to build on this substantial foundation is to incorporate it into computational modeling of the circuit, which can generate mechanistic predictions about its implications.

In this vein, we will build on past modeling work \cite{daun2011} to consider a model structure based on the experimentally derived interactions within the overall CPG circuit underlying the rhythmic movement of a single stick insect leg. 
We focus on the experimentally observed activation pattern of the motoneurons driving the six primary muscle groups of the three limb segments during forward stepping in single-leg insect preparations, leaving the interaction of limbs and the associated changes in neural outputs for future work.  The single-leg forward stepping pattern involves repeated sequences of movements that, in the context of locomotion on a level ground, consist of levation, or lifting, of the limb off of the ground; extension of the limb; a protraction that propagates the body forward; depression, or lowering of the limb; flexion of the limb that counters extension; and retraction that counters protraction.  Some of the associated neural outputs start or end at similar times, while others occur with characteristic phase shifts.  Roughly speaking, the time when protraction occurs can be used to define a swing phase of each step, with the period of retraction then corresponding to a stance phase \cite{fischer2001,bidaye2018}, although there may be some variations in this correspondence (cf. \cite{akay2004}).  

Previous modeling work showed that a CPG network based on the experimentally inferred set of neural influences could produce rhythmic activation consistent with the stick insect single-leg stepping pattern, at least transiently, and could serve as the basis for modeling interleg interactions \cite{daun2011}.  In this study, we more deeply investigate 
the patterns of sequential neuronal activation appropriate for driving the single-leg dynamics.
In the process of achieving this aim, we consider the following issues:

\begin{enumerate}
    \item how can the network connection strengths and intrinsic activation levels be tuned to  support sustained, functionally appropriate rhythm generation;
    \item within the resulting rhythms, what are the transition mechanisms -- known as escape and release \cite{skinner1994moa} -- through which activation switches from one to another antagonistic partner within each joint block, and what are the other dynamic features that yield the requisite phase relations between the active periods of different neuronal units;  
    \item what are the robustness properties of the resulting rhythms to parameter variations, and how do these relate to the dynamic features that underlie rhythm generation;
    \item how do parameter variations impact the period and phase durations of output patterns, and which forms of modulation, including neuromodulation affecting neuronal outputs or inputs, are predicted to be most effective at inducing functionally relevant pattern changes.

\end{enumerate}

The details of the models that we consider in pursuit of these aims are presented in Section \ref{sec:model}.  In Section \ref{sec:fundamental_concepts}, we review certain dynamical principles, including fast threshold modulation and transitions by escape and release, that are central to our subsequent analyses.  Most of our results appear in Section \ref{sec:escape_reduced}.   We find that the proposed neural circuitry and interactions can produce the fundamental activation pattern associated with single-leg stepping.  The parameter tuning that supports this form of dynamics features phase transitions entirely based on the escape mechanism, but with specific variations that we detail.  We then go on to expose and illustrate the implications for the robustness of rhythm generation to parameter variations that result from these dynamic mechanisms.  We find that there is significant variability in robustness with respect to different modulations, and we identify a way that neuromodulatory effects could tune the network's rhythmic output period and phase durations that may explain previous experimental observations \cite{gabriel2007}.  Finally, in Sections \ref{subsec:extension_inhibition}, \ref{subsec:extension_MN}, and \ref{sec:release}, we briefly address the impact of including certain additional features in the model and explain why release transitions are less well suited than escape transitions to produce a sustained rhythm.  We wrap up with a discussion in Section \ref{sec:discussion}.

%~~~~~~~~~~~~~~~~~~~~~~~~~~~~~~~~
\section{Mathematical Model}\label{sec:model}
In this section, we present the mathematical model, developed in \cite{daun2011}, that we use for the  neural \textit{circuit} within a middle or mesothoracic leg of stick insects.
\begin{figure}[h!]
    \centering
\includegraphics[width=\textwidth]{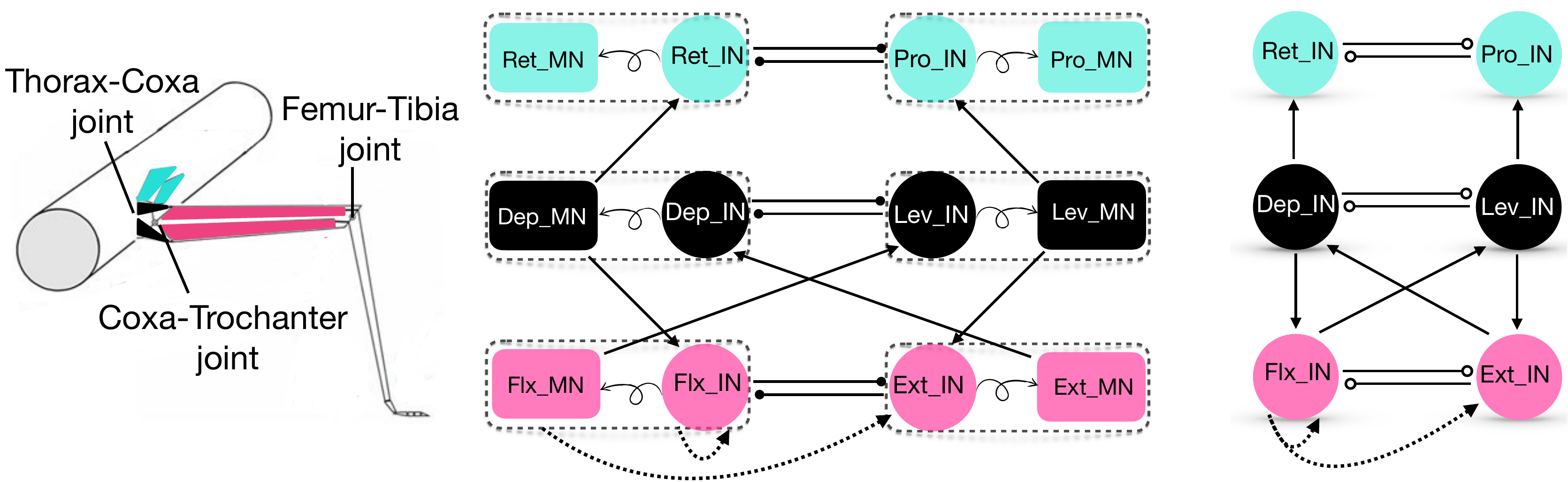}
    \caption{\textbf{The mesothoracic limb neural circuit.} 
    (Left) A schematic diagram of a stick insect’s middle leg with three main joints is shown.  Figure is adapted from Figure 6(b) in \cite{daun2011} with permission. (Middle) Each joint block consists of two interconnected IN-MN components that mutually inhibit each other ($\multimap$) and are represented by the same color. Each component, enclosed in a dashed box, includes a CPG IN pool and an MN pool. All MNs, except those in the top joint block, excite ($\rightarrow$) one or more adjacent joint blocks. Additionally,  Flx  excites both an adjacent joint block and its own joint block ($\dashrightarrow$).  The curly arrows represent an effective excitation (see Section \ref{sec:pair-CPG-MN}). 
    (Right) A simplified circuit is shown in which each component is represented solely by a CPG IN unit,  without MNs. }
    \label{fig:CIRCUIT} 
\end{figure}
As shown in Figure~\ref{fig:CIRCUIT}(middle panel), the mesothoracic limb is controlled by a neural circuit consisting of protraction (Pro)-retraction (Ret), levation (Lev)-depression (Dep), and extension (Ext)-flexion (Flx) \textit{joint blocks}. Each joint block, shown by a color, consists of two \textit{components}, each consisting of an interneuron (IN) unit and a motoneuron (MN) unit.  We use the term \textit{joint core} to refer to the pair of INs in the same joint block (the 3 joint cores appear in Figure~\ref{fig:CIRCUIT}, right panel).  Together, the INs in each joint core represent the central pattern generator (CPG) unit responsible for producing rhythmic output that drives the alternating activation of an antagonist muscle pair associated with one of the limb's joints, such as extensor and flexor muscles in the case of Ext-Flx.

We will consider certain variations on the  model for the mesothoracic limb circuit.  In the most complicated model (Figure~\ref{fig:CIRCUIT}, left panel), as proposed in \cite{daun2011}, the MN unit in each component of a joint block is inhibited by the IN unit of the \textit{antagonistic} component in the same joint block. For example, the MN unit in the Pro component within the Pro-Ret joint block is inhibited by the IN unit  of the Ret component.  This scheme ensures that the activation of an IN promotes the activation of the MN in its same component while respecting Dale's law.  In most of this paper, however, for ease of analysis and other reasons (see Section \ref{subsec:extension_MN} for more details), we omit the MNs from the model and consider an IN-only version (Figure~\ref{fig:CIRCUIT}, right panel).  Here, each joint block consists only of its core, and the INs in the same joint block both inhibit each other and serve (in some cases) as a source of excitation to other joint blocks, as discussed below in Section~\ref{sec:circuit-CPG-MN}.  While this reduction ignores Dale's law, it more parsimoniously provides the same effective coupling scheme as the model with MNs.

In the following sections, we first present the mathematical model for a single component (e.g., the Pro IN-MN component) in Section~\ref{sec:single-IN-MN}. Next, we describe the coupling within each joint block, including the inhibition between INs in the joint core, in Section~\ref{sec:pair-CPG-MN}. Finally, in Section~\ref{sec:circuit-CPG-MN}, we describe the coupling between joint blocks. 

%~~~~~~~~~~~~~~~~~~~~~~~~~~~~~~~~
\subsection{Single IN-MN component}\label{sec:single-IN-MN}
The mesothoracic limb circuit model comprises six IN-MN components. In this section, we present the model for a single component, which consists of an inhibitory IN unit and an excitatory MN unit, shown by dashed rectangles in Figure~\ref{fig:CIRCUIT} (middle panel). Each neuronal component is described using a three-dimensional system of ODEs, as follows:
\begin{subequations}\label{eq:single_CPG_MN}
\begin{align}
&\dot v=  F(v,h) = -\frac{1}{C_m}\left\{\Iin(v,h)  + \Iappin(v) + I_{Leak}(v)\right\},\label{eq:single_v}\\
&\dot h = \epsilon\; G(v,h) = \frac{\epsilon}{\tau_h(v)} \left(\hinfin(v) - h\right),\label{eq:single_h}\\
&\dot \w= K(\w) =  -\frac{1}{C_m}\left\{\Imn(\w) + I_{Leak}(\w)\right\},\label{eq:single_w}
\end{align}
\end{subequations}
where $(v,h,\w)^\top$ denotes the voltage and slow inactivation of the persistent sodium current of the interneuron and the voltage of the motoneuron within the component. 
The superscripts $^{IN}$ and $^{MN}$ stand for interneuron and motoneuron, respectively. 
Note that the MNs are decoupled from the INs of their own component and are controlled by the INs of the opposite component within the same joint block.  
Specifically, the curly arrow between the IN and MN of a component in Figure~\ref{fig:CIRCUIT} denotes ``effective" excitatory coupling between the IN and MN units within each component. That is, since the opposite IN within a joint block inhibits both the IN and MN within a component, and the INs inhibit each other, the activation of an IN effectively ``excites" its partner MN.  (See 
Section~\ref{sec:pair-CPG-MN} below for more details.)  In the simpler version of the model, for each component, we omit the MN and hence equation (\ref{eq:single_w}), while equations (\ref{eq:single_v}), (\ref{eq:single_h}) remain unchanged.

In equation (\ref{eq:single_v}), $C_m$ is the unit's membrane capacitance.
The persistent sodium current ($I_{NaP}$), applied current ($\Iappin$) such as that induced by a top-down drive from cortex, and leakage current ($I_{Leak}$) are given by the following equations.
\begin{subequations}
\begin{align}
&\Iin(v,h) = g_{NaP}\; m_{\infty}(v)\; h\; (v-E_{NaP})\label{eq:I_Nap_IN}\\
&\Imn(\w) = g_{NaP}\; m_{\infty}(\w)\; \hinfmn(\w)\;(\w-E_{NaP})\label{eq:I_Nap_MN}\\
&\Iappin(v) = g_{ton}\;(v-\Eexc) \label{eq:I_app_in_mn}\\
&I_{Leak}(x) = g_{Leak}\;(x - E_{Leak}),\label{eq:I_leak}
\end{align}
\end{subequations}
where the steady state functions  $m_{\infty}$ and $h_{\infty}$ are  sigmoidal and given by
\begin{subequations}
\begin{align}
&m_{\infty}(v) =\dfrac{\sigma_m}{1+\exp(v-\theta_m)} \label{eq:m_infty} \\
&\hinfin(v)=\dfrac{\sighin}{1+\exp(v-\thethin)} \label{eq:h_infty_IN}\\
&\hinfmn(w) = \dfrac{\sighmn}{1+\exp(w-\thethmn)} \label{eq:h_infty_MN}
\end{align}
\end{subequations}
and the time scale $\tau_h$ is 
\begin{align} \label{eq:tau_h} 
\tau_h(v) =\dfrac{1}{\cosh\Big(\frac{v -\theta_{\tau}}{2\sigma_{\tau}}\Big)}.
\end{align}
This formulation derives from earlier general work on CPGs \cite{daun2009}, which was adapted for modeling the non-spiking neurons \cite{buschges1995} in the stick insect locomotor CPG \cite{daun2012,dauntoth,codianni2020}. 
The default values of the parameters associated with each IN-MN component are detailed in Table~\ref{table:params}.  We assume that these parameters are identical across all six components, except for the constant conductance parameter $g_{ton}$ of the applied current terms (see equation~\eqref{eq:I_app_in_mn}), which varies across the components. The specific values of $g_{ton}$ are provided in Table~\ref{table:params_intrinsic}. These parameters are chosen to ensure that equation~\eqref{eq:single_CPG_MN} admits a stable steady state, consistent with the expectation that individual components within stick insect locomotor CPGs do not oscillate (i.e., alternate rhythmically between spiking phases and silent phases)  in isolation \cite{daun2011}.

%~~~~~~~~~~~~~~~~~~~~~~~~~~~~~~~~
\subsection{Inhibitory coupling between antagonistic components within a joint block} 
\label{sec:pair-CPG-MN} 
In this section, we describe the inhibitory synaptic coupling between the two components within each joint block. 
In the full model, the INs of each component inhibit both the INs and MNs of the adjacent component.  
These inhibitory currents are introduced as additional terms added on the right hand sides of the voltage dynamics of the INs and MNs in equations~\eqref{eq:single_v}, \eqref{eq:single_w}, represented by the following expressions:
\begin{subequations}\label{eq:I_syn}
\begin{align}
& \Isynin(v_i,v_j) =  \gsynin \;s_{\infty}(v_{j}) (v_i-\Einh) \label{eq:I_syn_IN} \\
& \Isynmn(w_i,v_j) = \gsynmn\; s_{\infty}(v_{j}) (w_i-\Einh).\label{eq:I_syn_MN} 
\end{align}
\end{subequations}
For example, the inhibitory currents from Ret-IN to Pro-IN and to Pro-MN are given by $\Isynin (v_1,v_2)$ and $\Isynmn (w_1,v_2)$, respectively,  where index 1 corresponds to the  Pro units (the post-synaptic neurons) and index 2 corresponds to the Ret-IN unit (the pre-synaptic interneuron). 

In equation \eqref{eq:I_syn}, the $g$ terms denote maximal conductances that would arise if the channels for these currents were completely open or activated. The $s$ terms are fractions of the channels that are actually open.  
We assume that the synapses quickly reach their steady states. Therefore, we approximate the open channel fractions using their steady-state values and neglect their dynamics. In \eqref{eq:I_syn}, each $s_{\infty}(v_{j})$ represents this steady-state, modeled by the sigmoidal function
\begin{align}\label{eq:s_infty1} 
s_{\infty}(v_{j}) = \dfrac{\sigma_{inh}}{1+\exp(v_{j} -\theta_{inh})}. 
\end{align}
  
All  parameters in equations~\eqref{eq:I_syn}-\eqref{eq:s_infty1} are fixed as specified in Table~\ref{table:params}, except for the inhibitory coupling strengths between the IN units of each joint block. These coupling strengths are denoted by  $\gsynin$, where $j\multimap i$ represents the inhibition from pre-synaptic IN $j$ to post-synaptic IN $i$, and their specific values are provided in Table~\ref{table:params_inh_coupling}. 
With these values, each joint core does not oscillate on its own and the circuit requires interaction between joint blocks to generate rhythmic activity, although each joint core does shift into an oscillatory mode with a modest increase in $g_{ton}$ in equation (\ref{eq:I_app_in_mn}) for both components, as seen in oscillation-inducing experimental manipulations (bath application of the muscarinic agonist pilocarpine \cite{buschges_etal_1995}; see Section \ref{sec:escape_reduced}). 

Importantly, in the reduced model that omits MNs on which we focus for most of our analysis, we can simply drop equation (\ref{eq:I_syn_MN}).  In the next section, we detail the excitatory coupling between joint blocks that induces oscillatory behavior, in the presence or absence of MNs.

%~~~~~~~~~~~~~~~~~~~~~~~~~~~~~~~~
\subsection{Synaptic coupling between  joint blocks}\label{sec:circuit-CPG-MN}

In this section, we describe the excitatory coupling between adjacent joint blocks. 
Which possible connections are included derives from a detailed influence diagram  worked out experimentally (summarized in \cite{bidaye2018}).  As shown in  Figure~\ref{fig:CIRCUIT}(middle panel), there is unidirectional coupling from the Lev-Dep joint block to the Pro-Ret joint (with no feedback from the Pro-Ret joint to the rest of the circuit), via the connections
 Lev $\rightarrow$ Pro and Dep $\rightarrow$ Ret. 
The Lev-Dep joint block also excites the Ext-Flx joint block and is excited by it in return:  Lev $\rightarrow$ Ext and Dep $\rightarrow$ Flx in one direction, and Ext $\rightarrow$ Dep and Flx $\rightarrow$ Lev in the other. 
Note that while Lev and Dep excite ipsilateral INs, Ext and Flx excite contralateral INs. Additionally, Flx excites INs within its own joint block: Flx $\rightarrow$ Flx and Flx $\rightarrow$ Ext.

Biologically, the presynaptic neurons for these excitatory interactions are likely either MNs or dedicated excitatory interneurons, yet in the reduced model on which we focus most of our work, we achieve the same connectivity with direct IN $\rightarrow$ IN connections.  In both cases, the excitatory synaptic currents are implemented by  incorporating terms of following form into the voltage dynamics of the INs in equation~\eqref{eq:single_v}:
\begin{align}
& \Isynexc(v_i,x_j) =  \gexji \;\tilde{s}_{\infty}(x_{j}) \;(v_i-\Eexc). \label{eq:I_syn_exc} 
\end{align}
Here, $\gexji$ denotes the maximal conductance of the excitatory coupling from the $j$ component to the IN unit in component $i$, and $\tilde{s}_{\infty}$ represents the steady-state open fraction of synaptic channels, modeled by the sigmoidal function
\begin{align}\label{eq:s_infty2} 
\tilde{s}_{\infty}(x_{j}) = \dfrac{\sigma_{exc}}{1+\exp(x_{j} -\theta_{exc})},
\end{align}
where $\sigma_{exc}$ and $\theta_{exc}$ are fixed parameters, as provided in Table~\ref{table:params}.   The term $x_j$ in equations (\ref{eq:I_syn_exc}), (\ref{eq:s_infty2}) denotes the voltage of the presynaptic neuron, which is $v_j$ in the reduced model and $w_j$ if MNs are included.
The values of the $\gexji$ parameters are given in Table~\ref{table:params_exc_coupling}.

Interestingly, the experimentally derived influence diagram is ambiguous in the sense that a  positive influence of one component on another in a different joint block could be effected either via a direct excitatory coupling or via an inhibition of the antagonistic IN within the target component's joint core, or by some combination of the two.  To allow for the possibility of an inhibitory influence mechanism, in some of our simulations, the excitatory coupling between joint blocks is paired with an inhibitory coupling such that, if one component excites another in a different joint block, then it also inhibits the antagonistic IN of that other joint block. For example, in this version of the model, if Lev excites Pro (Lev $\rightarrow$ Pro), then it also inhibits Ret (Lev $\multimapdot$ Ret). These inhibitory connections are described by
\begin{align}
 \Isyninh(v_i,x_j) = \ginji \; \hat{s}_{\infty}(x_{j}) \;(v_i-\Einh)\label{eq:I_syn_inh}, 
\end{align}
where we again represent the overall synaptic conductance as the product of a maximal conductance term and an open fraction term
\begin{align}\label{eq:s_infty3} 
\hat s_{\infty}(x_{j}) = \dfrac{\hat\sigma_{inh}}{1+\exp(x_{j} -\hat\theta_{inh})},
\end{align}
and where once again, the $x_j$ may be $v_j$ or $w_j$ depending on the model under consideration.
The values of  $\hat\sigma_{inh}, \hat\theta_{inh}$ and the $\ginji$ parameters are given in Tables~\ref{table:params_inh_exc_coupling} and~\ref{table:params_inh_coupling}, respectively. 
In Section~\ref{subsec:extension_inhibition}, we will 
discuss how this inhibitory coupling impacts circuit dynamics.

%~~~~~~~~~~~~~~~~~~~~~~~~~~~~~~~~
\section{Fundamental concepts for model analysis} \label{sec:fundamental_concepts}
In this section, we present a general overview of fast-slow systems and transitions between silent and active states characterized by escape and release phenomena. 
We also discuss three key mechanisms for silent-active transitions  arising from time-dependent excitatory inputs: fast threshold modulation, early excitation, and ghost mechanisms.
Readers already familiar with these concepts may skip ahead to Section~\ref{sec:escape_reduced}. 

\subsection{Fast-slow decomposition}\label{subsec:fast-slow}
Note that the neuron model equations (\ref{eq:single_v}),(\ref{eq:single_h}) take the form
\begin{subequations}
\label{eq:slow_fast}
\begin{align}
&\dot v=  F(v,h; \sexc, \sinhb) \\
&\dot h = \epsilon\; G(v,h) 
\end{align}
\end{subequations}
where $\sexc, \sinhb$ denote the total conductance of excitatory and inhibitory input, respectively, to a neuron.  Since we take $0 < \epsilon \ll 1$, reflecting the slow inactivation of the persistent sodium current, and $F, G$ are both ${\cal O}(1)$ with respect to $\epsilon$, these equations comprise a simple fast-slow system, with $v$ and $h$ evolving on widely disparate timescales.  

We have one copy of these model equations per interneuron unit  in our network.
Despite the overall high-dimensionality of the full network model, we can proceed by following a classic approach to studying coupled systems of model neurons, each represented by its own system of conductance-based differential equations such as (\ref{eq:single_v}),(\ref{eq:single_h}) and collectively interacting through synapses, which is to consider a projection of the network dynamics to a collection of two-dimensional phase spaces, one per neuron.
Since $v, h$ in (\ref{eq:slow_fast}) are scalar, our phase spaces will be two-dimensional, and the dynamics within each, if $\sexc, \sinhb$ are held fixed, can be analyzed using a singular limit analysis based on a simple fast-slow decomposition.

\begin{figure}[h!]
\centering
\includegraphics[width=.8\textwidth]{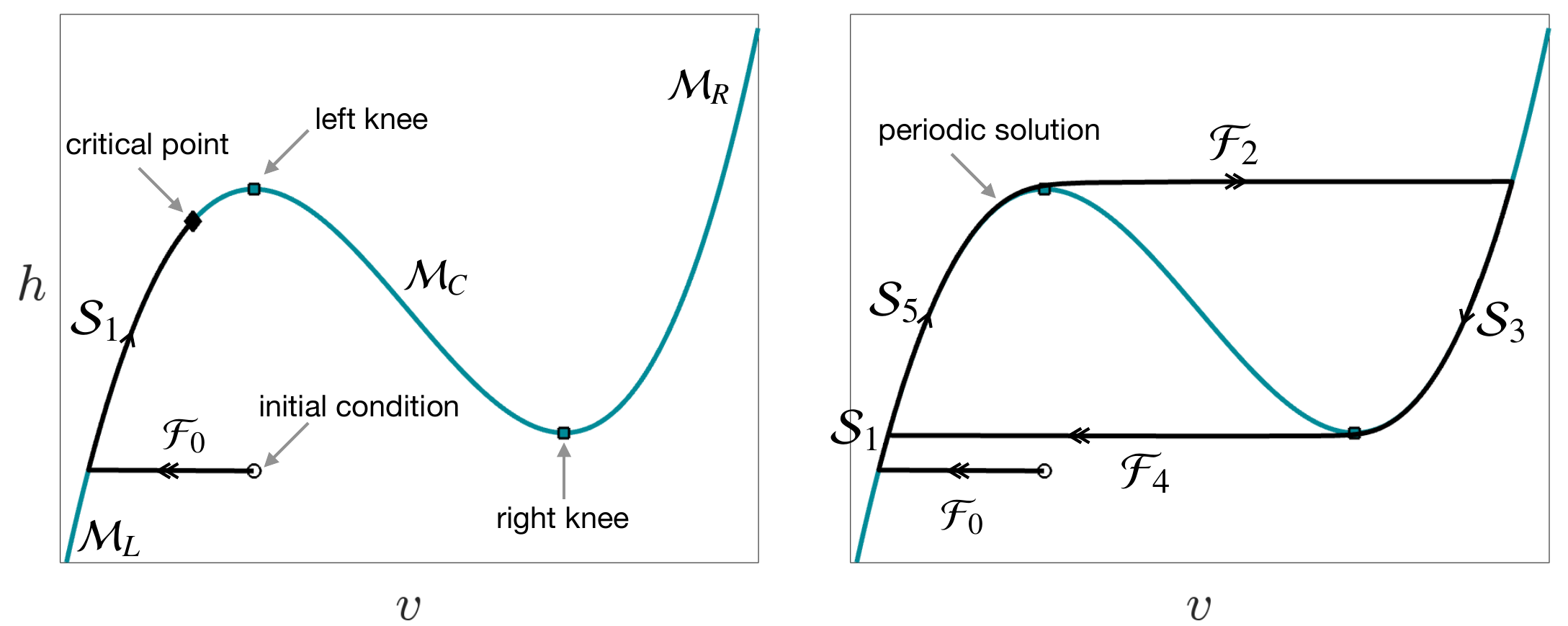}
    \caption{\textbf{Critical manifold and singular solutions of a fast-slow system.} The left panel shows a teal cubic-shaped critical manifold along with a black singular solution of a fast-slow system with a stable critical point. The right panel illustrates a system with no stable critical points, featuring a singular limit cycle. Singular solutions consist of the union of fast jumps alternating with slow flow orbits, represented as ${\cal F}_0 \cup {\cal S}_1$ on the left panel and ${\cal F}_0 \cup {\cal S}_1 \cup {\cal F}_2 \cup {\cal S}_3 \cup {\cal F}_4 \cup {\cal S}_5$ on the right panel. On the left panel, the singular solution converges to the stable critical point, while on the right panel, it converges to a limit cycle ${\cal F}_2 \cup {\cal S}_3 \cup {\cal F}_4 \cup {\cal S}_5$. For more details, see Section~\ref{subsec:fast-slow}.}
    \label{fig:fast-slow} 
\end{figure}

Specifically, for fixed $(\sexc,\sinhb)$, define ${\cal M} = \{ (v,h) : F(v,h;\sexc, \sinhb)=0 \}$; although ${\cal M}$ is called the critical manifold in geometric singular perturbation theory, in our planar case it is simply the $v$-nullcline.
For any initial condition $(v_0,h_0) \notin {\cal M}$, we approximate the solution of system (\ref{eq:slow_fast}) by the solution $(v(t),h_0)$ of the fast subsystem obtained with $v(0)=v_0$ and $\epsilon=0$. 
In our case, we assume that for each relevant $(\sexc, \sinhb)$ pair, the manifold ${\cal M}$ is given by the graph over $v$ of a smooth function $h(v;\sexc, \sinhb)$ that is cubic-shaped in the $(v,h)$ plane.  Equivalently, we have ${\cal M} = {\cal M}_L \cup {\cal M}_C \cup {\cal M}_R$, the union of three branches that are, respectively, the graphs of smooth functions $v_L(h; \sexc, \sinhb),  v_C(h; \sexc, \sinhb)$, and
$v_R(h; \sexc, \sinhb)$.  
We assume that on their interiors, the left and right branches ${\cal M}_L, {\cal M}_R$ are attracting for the fast subsystem, that the interior of the central branch ${\cal M}_C$ is repelling, and that the outer branches meet the central branch at points where $h'(v)=0$.  These points represent fold bifurcations in the set of equilibria of the fast subsystem, with $h$ treated as a bifurcation parameter.  We follow standard terminology and refer to these points as knees, denoted by $(v_{LK}(\sexc, \sinhb),h_{LK}(\sexc, \sinhb))$ for the left knee and $(v_{RK}(\sexc, \sinhb), h_{RK}(\sexc, \sinhb))$ for the right knee.  Note that while various quantities such as $v_L, v_{LK}, \ldots$ depend on $(\sexc,\sinhb)$, we will henceforth omit mention of this dependence except when it is important for clarity.

Now, returning to the fast-slow decomposition:  Under the assumptions we have given, for $(v_0,h_0) \notin {\cal M}_C$, it follows that  $(v(t),h_0) \to {\cal M}_L \cup {\cal M}_R$ as $t \to \infty$.  We refer to the orbit ${\cal F}_0 := \{ (v(t),h_0) : t \in [0,\infty) \}$ as a fast jump. 
Suppose, for example, that $v(t) \to v_L(h_0)$ as $t \to \infty$.  We derive a second solution component from the flow of the slow subsystem 
$h' = G(v_L(h),h)$, where $'$ denotes the rescaled time $\tau = \epsilon t$; we call this solution ${\cal S}_1 = \{ (v_L(h),h) \}$.   
This component is well-defined as long as $v_L(h)$ is.  If this orbit reaches the left knee, then $v_L(h)$ is lost and we consider a second fast jump, say ${\cal F}_2:= \{ (v(t),h_{LK}) \}$ from $(v_{LK},h_{LK})$ to $(v_R(h_{LK}),h_{LK})$.  This jump is followed by a subsequent slow flow that yields an orbit ${\cal S}_3 := \{ (v_R(h),h) \}$ on ${\cal M}_R$.  Proceeding in this way, we define a singular solution, consisting of the union of fast jumps alternating with slow flow orbits ${\cal F}_0 \cup {\cal S}_1 \cup {\cal F}_2 \cup {\cal S}_3 \cup \ldots$.  Such a singular solution may converge to a periodic oscillation (e.g., in Figure \ref{fig:fast-slow}, right panel, ${\cal F}_2 \cup {\cal S}_3 \cup {\cal F}_4 \cup {\cal S}_5$) or may terminate with an orbit of the slow flow converging as $\tau \to \infty$ to a full system critical point, $(v_{CP}(\sexc, \sinhb), h_{CP}(\sexc, \sinhb))$, where $F(v_{CP},h_{CP}) = G(v_{CP},h_{CP})=0$ (e.g., in Figure \ref{fig:fast-slow}, left panel). 
Existing theory shows that for these types of models, actual solutions exist that are $C^1$ $O(\epsilon)$-close to such a singular solution away from the knees, with a weaker form of proximity preserved even at the knees \cite{fenichel1979,mkkr,jones1995}.

What about the fact that the arguments $\sexc, \sinhb$ for each neuron actually vary in time and depend on the voltages of other neurons?  We incorporate effects of changes in synaptic conductances into the fast-slow decomposition through their effects on the location of each $v$-nullcline, since these factors appear only in the $v$-equation in (\ref{eq:slow_fast}). 
That is, changes in $\sexc, \sinhb$ terms in our model will only occur during fast jumps in which a pre-synaptic or source neuron's voltage transits between ${\cal M}_L$ and ${\cal M}_R$, in either direction.  Thus, we keep track of the fast-slow dynamics in the phase planes of all of our model neurons, and when one makes such a fast jump, we correspondingly adjust the $v$-nullclines of all of the post-synaptic neurons that it targets \cite{rubin2002,ermterm}.  Thus, in a coupled neuron model, the projection of a solution to the phase plane for one neuron may include  not only fast jumps between left and right $v$-nullcline braches and slow orbits, but also fast jumps between left  or right branches of different $v$-nullclines, say from $(v_L(h;\sexc,\sinhb),h)$ to $(v_L(h;\tilde{g}_{exc},\tilde{g}_{inh}),h)$, where at least one of $(\tilde{g}_{exc},\tilde{g}_{inh})$ differs from the corresponding element of $(\sexc,\sinhb)$.

\subsection{Escape and release transitions}\label{subsec:escape_release}

\begin{figure}[h!]
\centering
\includegraphics[width=1\textwidth]{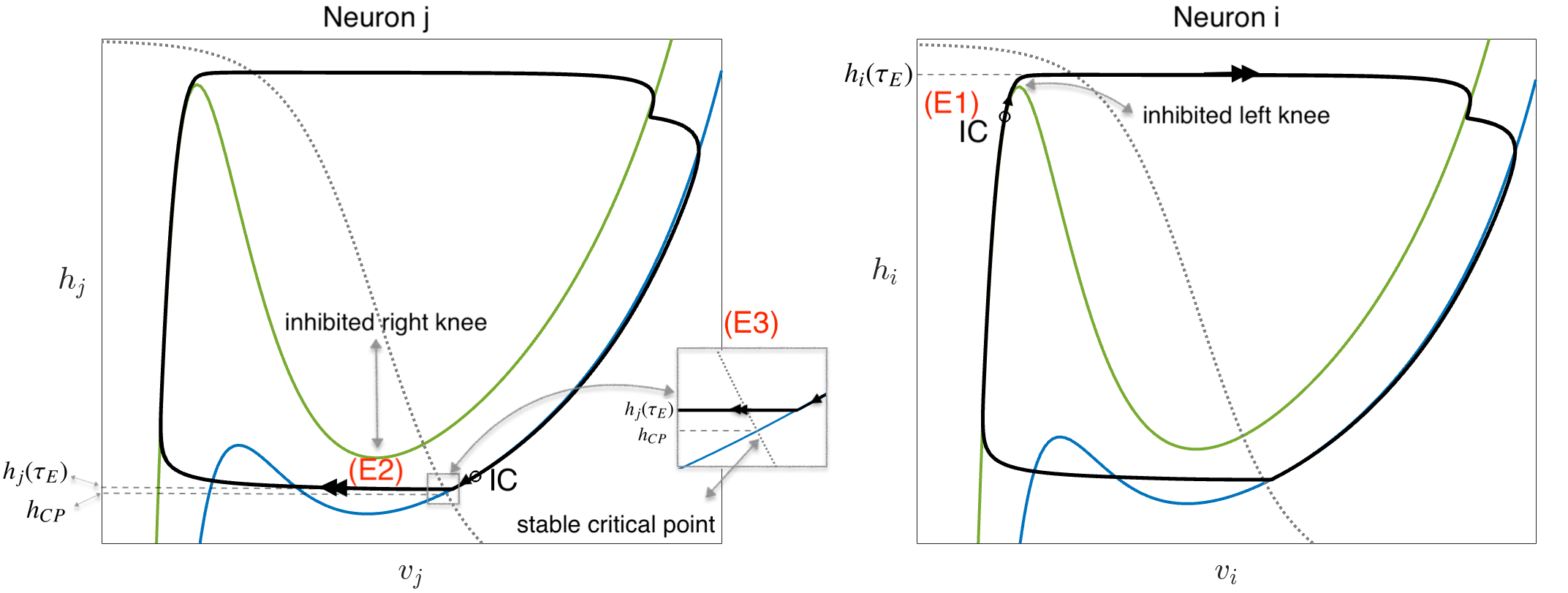}
    \caption{\textbf{Generation of periodic solutions through escape.} The $(v,h)$ phase planes of two mutually inhibited model neurons, each as described in equation~\eqref{eq:slow_fast}, are shown. Two cubic-shaped $v$-nullclines are depicted: the green nullcline corresponds to the inhibited state of a neuron, while the blue nullcline represents the uninhibited state. The dotted gray curves denote the $h$-nullclines, and IC indicates the initial conditions for each neuron. 
    Conditions ($E1$), ($E2$), and ($E3$),   which describe the requirements for the escape mechanism, are detailed in Section~\ref{subsec:escape_release}.
    For example, on the right panel, there is no critical point on the left branch of inhibited green v-nullcline. Therefore, condition ($E1$) holds and the solution starting from IC reaches the left knee and  jumps to the right branch at time $\tau_E$. That is, neuron $i$ escapes from inhibition and activates.  On the left panel, we see that $h_j(\tau_E)$ is below the right knee of the inhibited green $v$-nullcline and above the stable critical point $(v_{CP},h_{CP})$. Therefore, conditions ($E2$) and ($E3$) are satisfied, such that neuron $j$ could not jump down on its own but does so when the escape of neuron $i$ causes it  to become inhibited. }
    \label{fig:escape} 
\end{figure}

As we have just discussed, for a model neuron governed by equations 
(\ref{eq:single_v}),(\ref{eq:single_h}), the neuron's intrinsic dynamics is determined by the intersections between the $v$- and $h$-nullclines.  In the $\epsilon \to 0$ singular limit, if these curves intersect once, then the resulting critical point is asymptotically stable if it lies on ${\cal M}_L \cup {\cal M}_R$ and is unstable if it lies on ${\cal M}_C$; in the latter case, the singular attractor is periodic.  For $0 < \epsilon \ll 1$, the stability transitions move away from the knees of the $v$-nullcline but remain $O(\epsilon)$-close.  

When neurons are coupled synaptically, the attractors depend on the intersections of the neurons' families of $v$-nullclines. JR{In this subsection, we will focus on inhibitory coupling, which could be the reciprocal inhibition within a joint core, as given in equation (\ref{eq:I_syn_IN}), or the inter-joint coupling specified in equation (\ref{eq:I_syn_inh}).}
Note that $s_{\infty}(v) \in (0,1)$, and hence for neuron $i$ inhibited by neuron $j$, the $v$-nullcline family can be parameterized by $\sinhb \in (0,\gsynin)$.

In the singular limit, however, since changes in $v$ occur on the fast timescale,  we have $\sinhb \in \{0,\gsynin\}$, and hence we only have to consider a pair of $v$-nullclines rather than a continuum of them.  Since we will ignore excitation for the time being, we will refer to $F(v,h;g_{inh}), G(v,h;g_{inh})$, and so on, omitting the $g_{exc}$ argument.

In this setting, two canonical cases of interactions between model neurons associated with anti-phase activity  have been distinguished \cite{wang1992,skinner1994moa}:  {\em escape} and {\em release}.
In the escape regime, a key assumption is
\[
(E1) \qquad G(v_L(h),h;\gsynin) \neq 0
\]
for all $h < h_{LK}(\gsynin)$. 
That is, when neuron $i$ is inhibited by neuron $j$, there are no critical points on the left branch of $v$-nullcline for neuron $i$.
Thus, despite the inhibition to neuron $i$, any slow orbit 
$\{ (v_L(h_i(\tau);\gsynin),h_i(\tau)) \}$ 
will reach $(v_{LK}(\gsynin),h_{LK}(\gsynin))$,  after which a transition to a fast jump up to the right branch ${\cal M}_R$ of the inhibited $v$-nullcline for neuron $i$ will follow (e.g., in Figure~\ref{fig:escape}(right panel)).  That is, neuron $i$ will {\em escape} from inhibition and activate, say at time $\tau = \tau_E$.
Going one step further, under the additional condition 
\[
(E2) \qquad  h_j(\tau_E) < h_{RK}(\gsyninij),
\] 
 the inhibition from neuron $i$ to neuron $j$ associated with the activation of neuron $i$ will cause neuron $j$ to make a fast jump down to the silent phase (e.g., Figure~\ref{fig:escape}(left panel)).
If conditions $(E1), (E2)$ hold for both neurons (i.e., with $i, j$ switched as needed and with $\tau_E$ replaced by the time of each subsequent escape event), 
then anti-phase, alternating activity results.
We can ensure that the activation switches continue to occur due to escape by including a third condition,
\[
(E3) \qquad \{ (v,h) : F(v,h;0) = 0 = G(v,h;0) \} = (v_{CP}(0),h_{CP}(0)) 
\]
where $v_{CP}(0) = v_R(h_{CP}(0);0)$ and $h_{CP}(0) \in (h_{RK}(0),h_{RK}(\gsyninij))$ (e.g., in Figure~\ref{fig:escape}(left panel - zoomed box)). 
Condition (E3) implies that 
$(v_{CP}(0),h_{CP}(0))$ is an asymptotically stable critical point for the dynamics of neuron $j$, with $\sinhb=0$ fixed, and hence 
under condition (E3), neuron $j$ cannot leave the active phase before neuron $i$ can escape.

\begin{figure}[h!]
\centering
\includegraphics[width=1\textwidth]{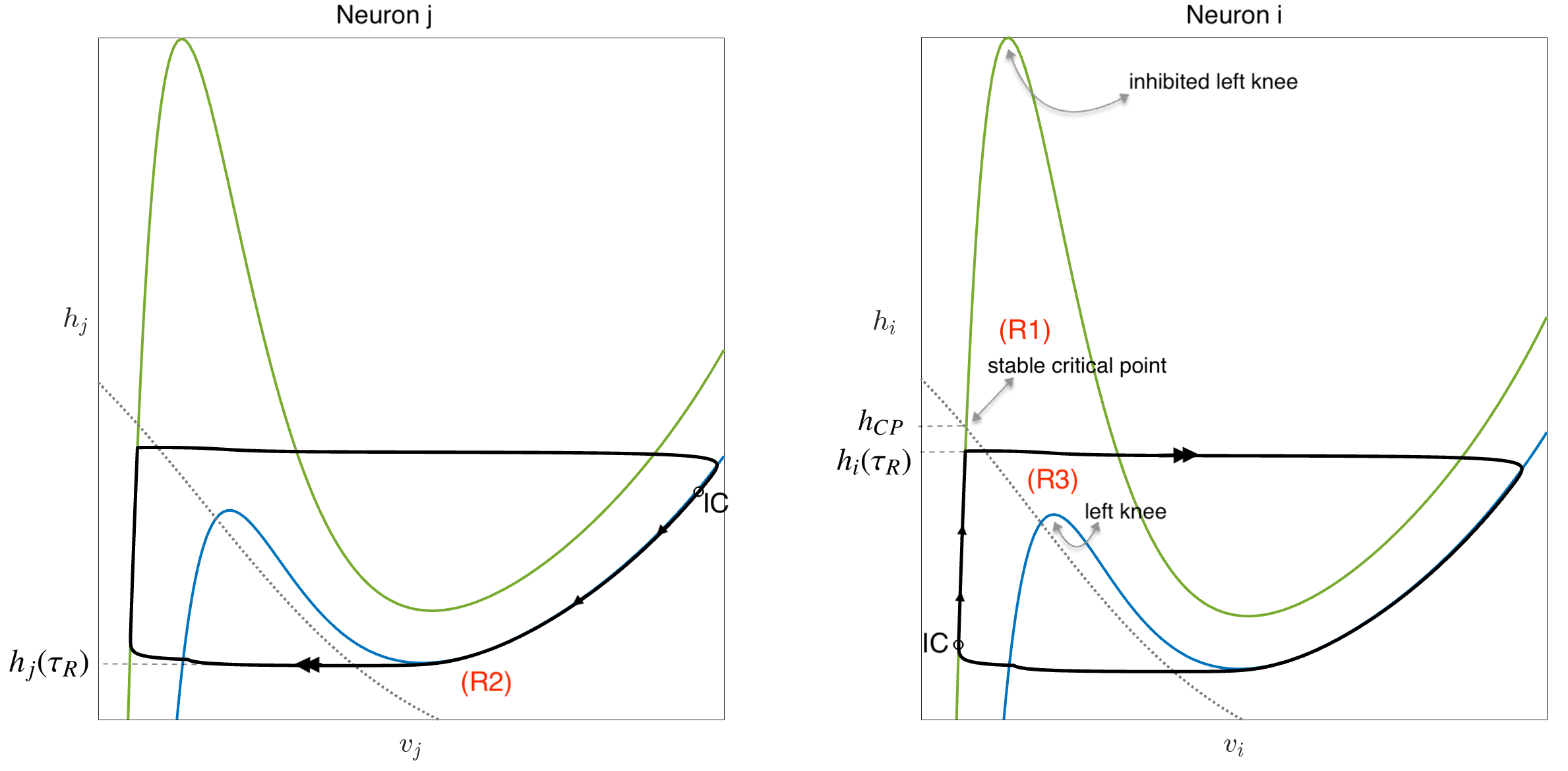}
\caption{\textbf{Generation of periodic solutions through release.} The $(v,h)$ phase planes of two mutually inhibited model neurons, each as described in equation~\eqref{eq:slow_fast}, are shown. Two cubic-shaped $v$-nullclines are depicted in each: the green nullcline corresponds to the inhibited state of each neuron, while the blue nullcline represents the uninhibited state. The dotted gray curves denote the $h$-nullclines, and IC indicates the initial conditions for each neuron. 
    Conditions ($R1$), ($R2$), and ($R3$), analogous to ($E3$), ($E1$), and ($E2$),     describe the requirements for the release mechanism and are detailed in Section~\ref{subsec:escape_release}.
    }
    \label{fig:release} 
\end{figure}

In the release regime, the analogue of assumption (E3) is
\[
(R1) \qquad \{ (v,h) : F(v,h;\gsynin) = 0 = G(v,h) \} = (v_{CP}(\gsynin),h_{CP}(\gsynin)) 
\]
where  $v_{CP}(\gsynin)=v_L(h_{CP}(\gsynin);\gsynin)$ and $h_{CP}(\gsynin) \in  
(h_{LK}(0),h_{LK}(\gsynin)).$
Under condition (R1), $(v_{CP}(\gsynin),h_{CP}(\gsynin))$  
is an asymptotically stable critical point for the dynamics of neuron $i$, for $\sinhb = \gsynin$ fixed, such that neuron $i$ cannot escape from the silent phase while inhibited by neuron $j$.
A second condition analogous to (E1), namely
\[
(R2) \qquad G(v_R(h),h;0) \neq 0, 
\]
for all $h$ such that $v_R(h)$ is defined, ensures that the slow orbit for neuron $j$ in the active phase will reach $(v_{RK}(0),h_{RK}(0))$ and transition to a fast jump down to $(v_L(h_{RK}(0);0),h_{RK}(0))$, say at time $\tau_R$.
Finally, since this jump down causes the inhibition from neuron $j$ to neuron $i$ to drop to 0, corresponding to a {\em release} of neuron $i$, the condition 
\[
(R3) \qquad h_i(\tau_R) > h_{LK}(0)
\]
will ensure that as a result, neuron $i$ undergoes fast activation by jumping up to the active phase.
If conditions $(R1)-(R3)$ hold for both neurons for all transitions (i.e., with $i, j$ switched as needed and with $\tau_R$ replaced by the time of each subsequent release event), then anti-phase, alternating activity, with transitions by release, will result.

 \subsection{Excitation-induced escape:   early excitation (EE), fast threshold modulation (FTM), and ghost of saddle-node (GSN) mechanisms}\label{sec:FTM-EE-Gh}

In the previous subsection, we discussed transitions between silent and active phases resulting from intrinsic dynamics and from effects of rapid changes in inhibition.  The models of interest in this work also include excitation, which in our modeling framework will also turn on and off quickly.
Relative to escape and release, the ways by which changes in excitation can contribute to phase changes in rhythmic systems have generally received less attention (but see \cite{somers1993,rubin2005}), and it will be important for our results to highlight certain of these mechanisms for the escape regime.

Recall that in the standard escape case (e.g., Figure \ref{fig:escape}), the silent neuron must be able to jump up to the active phase despite receiving inhibition (E1).  The active neuron must not jump down from the active phase on its own (E3), but must be forced to do so when it becomes inhibited.
If a silent neuron receives an excitatory input of strength $g_{exc}$ in this configuration, then this input may help it to activate.  Indeed, condition  (E1), or $G(v_L(h),h;0,g_{inh}) \neq 0$, no longer needs to hold once excitation is present; instead, we require
\[
(E1)_e \qquad G(v_L(h),h;g_{exc},g_{inh}) \neq 0
\]
for all $h < h_{LK}(g_{exc},g_{inh})$.
Similarly, we require
\[
(E2)_e \qquad h_j(\tau_E) < h_{RK}(g_{exc},g_{inh}),
\]
where $g_{exc}$ is the level of excitation (possibly 0) to the active neuron and $g_{inh}$ is the strength of inhibition it will receive when the silent neuron activates, and 
\[
(E3)_e \qquad \{ (v,h) : F(v,h;g_{exc},0) = 0 = G(v,h;g_{exc},0) \} = (v_{CP}(g_{exc},0),h_{CP}(g_{exc},0)), 
\]
with $g_{exc}, g_{inh}$ as above, and 
\[v_{CP}(g_{exc},0) = v_R(h_{CP}(g_{exc},0);g_{exc},0), \quad
h_{CP}(g_{exc},0) \in (h_{RK}(g_{exc},0),h_{RK}(g_{exc},g_{inh}).\]  
In fact, $(E2)_e$ and especially $(E3)_e$ are often effectively similar to $(E2)$ and $(E3)$, because the excitatory reversal potential $E_{exc}$ is near $v_{CP}$ over the relevant ranges of its arguments.  Hence, we will focus on $(E1)_e$ and how excitation can contribute to the initiation of escape.
 
Suppose that a neuron in the silent phase, at $(v_L(h_0;g_{exc},g_{inh}),h_0)$ for some $h_0$, experiences an abrupt increase in excitatory input, say to $g_{exc}+\Delta$ for $\Delta>0$.
According to analysis in the singular limit within the fast-slow decomposition framework, if $h_0 > h_{LK}(g_{exc}+\Delta,g_{inh})$, then the neuron will jump up to the active phase on the fast timescale and hence instantaneously on the slow timescale, while if $h_0 < h_{LK}(g_{exc}+\Delta,g_{inh})$, then the neuron will undergo a much smaller increase in $v$, jumping to $(v_L(h_0;g_{exc}+\Delta,g_{inh}),h_0)$ on the fast timescale.
Away from the singular limit, however, the hard distinction between these scenarios becomes blurred and additional factors come into play.  Based on these considerations, we will distinguish three key cases on the continuum of possible effects of excitatory inputs that contribute to escape, which we will call {\em early excitation, fast threshold modulation}, and {\em ghost transitions} and which have  distinct implications for the timing of phase transitions and for robustness with respect to parameter changes.

In general, we will assume that excitation increases by $\Delta>0$ when the neuron's trajectory is at $(v_L(h_0;g_{exc},g_{inh}),h_0)$ and that {$(E1)_e$ holds with $g_{exc}$ replaced by $g_{exc}+\Delta$ for all $h < h_{LK}(g_{exc}+\Delta,g_{inh})$,} 
such that there is no critical point on the $v$-nullcline for the elevated excitation level $g_{exc}+\Delta$.
In addition, let us first suppose that $h_0$ is sufficiently less than $h_{LK}(g_{exc}+\Delta,g_{inh})$.
We will refer to this case as {\em early excitation} (EE):  the neuron receives excitation, and with this excitation level, condition $(E1)_e$ 
ensures that the neuron will eventually transition from the silent to the active phase in the absence of subsequent inputs, but it will do so only after an excursion on the slow timescale that lasts for a positive duration of slow time, which we will denote as $\tau_{EE}>0$.
To cast EE in a fully rigorous framework, we would replace ``sufficiently less than'' with ``an $\mathcal{O}(1)$ distance below'', which would ensure that $\tau_{EE}$ is $\mathcal{O}(1)$.  In this case, the existence of an EE transition would be robust to $\mathcal{O}(\epsilon)$ parameter perturbations. We will not aim for this level of precision in our analysis, but we still note that EE will be structurally robust to parameter variations as long as $h_{LK}(g_{exc}+\Delta,g_{inh})-h_0$ is not too small.  The sensitivity of $\tau_{EE}$ to parameter variations can vary and depends on both $\Delta$ and on the product $(\partial v_L / \partial g_{exc})(\partial G / \partial v)$.

Next, we continue to suppose that excitation increases by $\Delta>0$ when the neuron's trajectory is at some point $(v_L(h_0;g_{exc},g_{inh}),h_0)$ and that $(E1)_e$  
holds.  We now impose two new assumptions on $h_0$, however.  First, we assume that
$h_0$ is sufficiently greater than than $h_{LK}(g_{exc}+\Delta,g_{inh})$.  Second, recall that $\{ h = h_{\infty}(v) \}$ defines the $h$-nullcline from equation (\ref{eq:single_h}).  We assume that for all $v \in (v_{L}(h_0;g_{exc},g_{inh}),v_L(h_0;g_{exc}+\Delta,g_{inh}))$, the quantity $h_{\infty}(v) - h_0$ is positive and bounded sufficiently away from zero.
In this case, the trajectory will immediately undergo a fast timescale excursion from its starting point to $(v_R(h_0;g_{exc}+\Delta,g_{inh}),h_0)$, which is instantaneous on the slow timescale.  This transition was named {\em fast threshold modulation} (FTM) in previous work that analyzed the phenomenon and its potential to induce rapid synchronization \cite{somers1993,rubin2002}. FTM transitions are highly robust to parameter variations, because small changes in $h_0$ will not affect the fact that when excitation increases by $\Delta$, the trajectory has $h_0 > h_{LK}(g_{exc}+\Delta,g_{inh})$ and also lies below and away from the $h$-nullcline, such that the excitation boost immediately results in a jump up to the active phase that is instant on the slow timescale.

Now, let us consider one of the situations not covered by these first two scenarios, still with $(E1)_e$.  
If $h_0 \approx h_{LK}(g_{exc}+\Delta,g_{inh})$, then, relative to FTM, the trajectory may experience a delayed passage near the left knee of the $v$-nullcline for the elevated excitation level, $(v_{LK}(h_{LK}(g_{exc}+\Delta,g_{inh})),h_{LK}(g_{exc}+\Delta,g_{inh}))$, as it transitions from the silent to the active phase \cite{mkkr}.  This delayed passage may follow a brief EE period or not, depending on the relative sizes of $h_0$ and $h_{LK}(g_{exc}+\Delta,g_{inh})$.  The duration of the transition delay itself is positive on the slow timescale although it goes to 0 with $\epsilon$.  It is thus significant relative to a pure FTM transition and insignificant relative to $\tau_{EE}$.  This transitional configuration, between FTM and EE, could easily be transformed into FTM or EE by small parameter variations, while quantitative solution properties in this regime may be sensitive to parameter changes in a similar way to $\tau_{EE}$ itself; correspondingly, we do not refer to it as a distinct case like EE and FTM.  

We note that both in this transitional scenario and in EE itself, although a slow excursion after the increase in excitation may be needed to allow the trajectory to leave the neighborhood of $(v_{LK}(h_{LK}(g_{exc}+\Delta,g_{inh})),h_{LK}(g_{exc}+\Delta,g_{inh}))$ and jump up to the active phase, the passage time is usually sped up by the fact that the trajectory lies below the $h$-nullcline.  As a result, away from the strict singular limit, $h$ initially increases during the passage near the left knee.  The increase in $h$ pulls the trajectory away from the knee, where $F=0$ in equation (\ref{eq:slow_fast}) and hence $\dot{v}$ is small, and  hence helps the fast timescale transition to the active phase to ensue.  

There is a final case, however, when an impactful increase in $h$ is prevented because $h_{\infty}(v)-h_0$ is too small over the relevant range of $v$ values, in violation of the second hypothesis of FTM.  Hence, after excitation increases,  a period occurs, possibly preceded by an EE segment, when the trajectory lies close to both the $v$- and $h$-nullclines. 
In this case, even if $h_0 > h_{LK}(g_{exc}+\Delta,g_{inh})$, the initial part of the transition has a duration $d \uparrow \infty$ as $\epsilon \downarrow 0$.  We refer to this form of delayed transition as a {\em ghost saddle-node} (GSN) mechanism.  This nomenclature refers to the fact that such a close proximity of the two nullclines naturally arises just beyond a saddle-node bifurcation in which two full system critical points come together and are lost;  when the critical points are present, no transition to the active phase can occur, and just beyond the bifurcation that destroys them, the {\em ghost} of their existence manifests in a slow transition due to the nullcline proximity \cite{izhikevich,hastings2018}.  The GSN mechanism is not generally robust to parameter variations, since small alterations in nullcline positions can have large effects on ghost transition times, regardless of whether or not a preliminary EE segment occurs.
 
%~~~~~~~~~~~~~~~~~~~~~~~~~~~~~~~~
\section{Generation of stepping rhythms in  insects' limbs:  transitions by escape}\label{sec:escape_reduced}

The stick insect locomotor CPG produces rhythmic activity that drives locomotion in the intact animal as well as a fictive locomotor rhythm, which emerges when the CPG circuit is isolated and hence feedback signals from stepping, ground contact, and so on are not present. 
Our first goal was to tune the parameters of the mesothoracic model to elicit an idealized 
time course of sequential unit activation that has been reported in the literature during the fictive locomotor rhythm \cite{buschges1994,fischer2001,akay2004,daun2009,bidaye2018}.
This rhythm features the following properties, which we henceforth call (P):
\begin{enumerate}
    \item The pattern is periodic in time\footnote{We do not expect to observe true periodicity in a neural recording. The idea is that periodic dynamics in the deterministic modeling framework that we consider will translate into robust, repetitive cycling in the presence of biological stochasticity.}.
    \item Within each joint core, the two interneuron units take turns being active.
    \item The activation onsets of the Ext and Lev units are sufficiently synchronized.
    \item The Pro unit's activation occurs approximately halfway through the active phase of Lev.
     \item The Lev unit's activation ends approximately halfway through the active phase of Pro.
    \item The ends of the active phases of the Ext and Pro units (equivalently the activation onsets of Flx and Ret) are sufficiently synchronized.
\end{enumerate}

We note that attaining these properties from the coupling architecture shown in Figure \ref{fig:CIRCUIT} is non-trivial. For example, when Lev activates, it excites both Pro and Ext; these two targets must nonetheless  activate at different times, while still deactivating together, to satisfy (P).

In the following, we first show results from the circuit depicted in Figure~\ref{fig:CIRCUIT} (right panel), where motoneurons are excluded, and the inter-joint excitatory couplings originate from an IN pool. These results include simulations obtained using XPPAUT \cite{xppbook} and MATLAB  and analysis based on the concepts and frameworks presented in Section \ref{sec:fundamental_concepts}. We explain the mechanisms underlying the rhythms generated by this circuit and discuss the robustness of its parameters. Subsequently, we extend the circuit by incorporating inter-joint  inhibitory coupling in Section~\ref{subsec:extension_inhibition} and motoneurons in Section~\ref{subsec:extension_MN}. We then evaluate how these extensions may influence the generation of a robust and ideal rhythm (P), identifying potential advantages and complications that they introduce.

\begin{figure}[h!]
    \centering
\includegraphics[width=.8\textwidth]{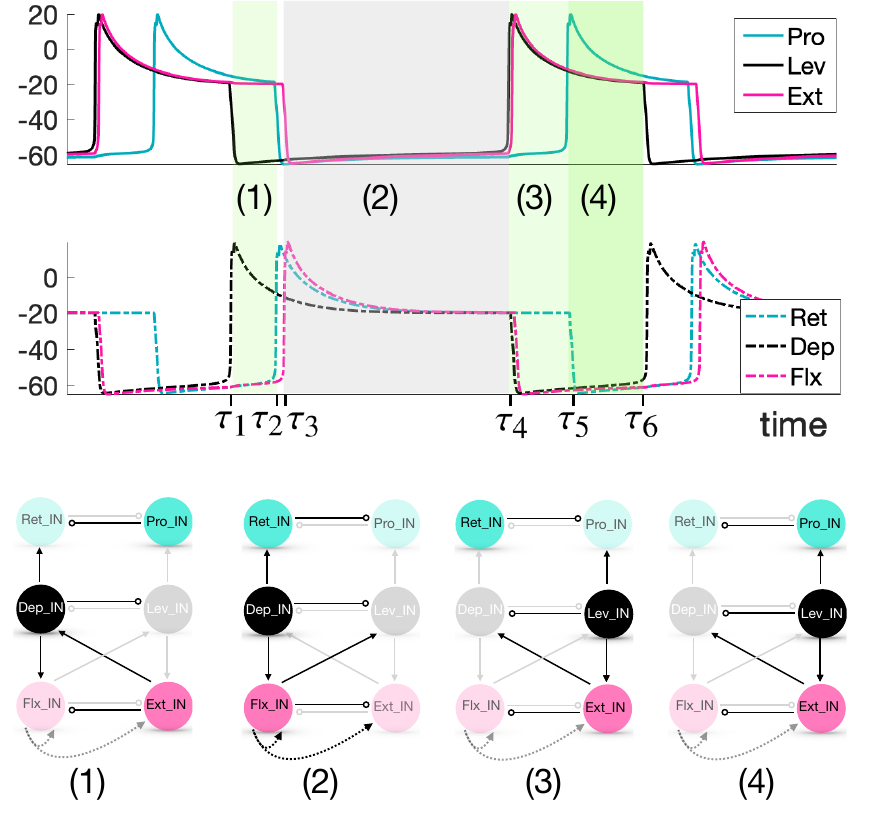}
    \caption{{\bf A baseline locomotor rhythm with timing properties (P).}
    Top panel: Voltage time courses of INs, matching properties observed in stick insect fictive locomotion \cite{buschges1994,fischer2001,akay2004,daun2009,bidaye2018}. The top set of traces show a set of INs that interact synergistically to achieve leg lift and advance (swing), while the bottom traces show their respective antagonists (for stance) in matching colors (Figure~\ref{fig:CIRCUIT}).  In the grey phase (2), the stance INs are active; in the darker green phase (4), the swing INs are active; and the pale green phases (1), (3) represent intermediate transitions.
     The model parameter values used are specified in the columns for ``Escape" or ``only IN, Escape" of Tables~\ref{table:params}-\ref{table:params_exc_coupling} in Section~\ref{Sec:appendix}.
     Bottom panel: The transition from active (bold color) to silent (pale color) phase or silent to active phase for each IN unit over one step cycle.  
     }
    \label{fig:time-series-IN} 
\end{figure}

\subsection{Escape-based rhythms with baseline parameter sets}

We consider a neural rhythm  generated by a mesothoracic limb locomotor CPG circuit composed of three coupled joint blocks, each including a joint core of two INs interacting by reciprocal inhibition. This rhythm  features multiple phases of prolonged activation of different INs, and since the inactive IN in each block is subject to inhibition from its active partner, either escape or release could be associated with each phase transition in this rhythm.
In practice, we found that producing a stable rhythm with the desired timing and phase relations required a tuning that featured phase transitions by escape.  The activation time courses of the six IN units in the circuit for a baseline parameter set (see Section \ref{Sec:appendix}), with transitions by escape, are shown in Figure \ref{fig:time-series-IN}.  Note that in an intact animal, activation of Lev drives  levator muscles that lift the leg, subsequent activation of Pro drives protractor muscles that move the leg forward, and finally activation of Dep drives depressor muscles that return the leg to the ground; thus, a locomotor swing phase can roughly be defined as the time from the start of Lev activation until the end of Pro activation, which is roughly 40$\%$ of the total cycle duration \cite{fischer2001,grabowska2012}. 
We next turn to phase plane analysis to explain how the features of this rhythm emerge and to lay the groundwork for understanding robustness of this rhythm to parameter variations.   

\subsection{Nullcline analysis of escape-based rhythms}
\begin{figure}[h!]
    \centering
\includegraphics[width=.25\textwidth]{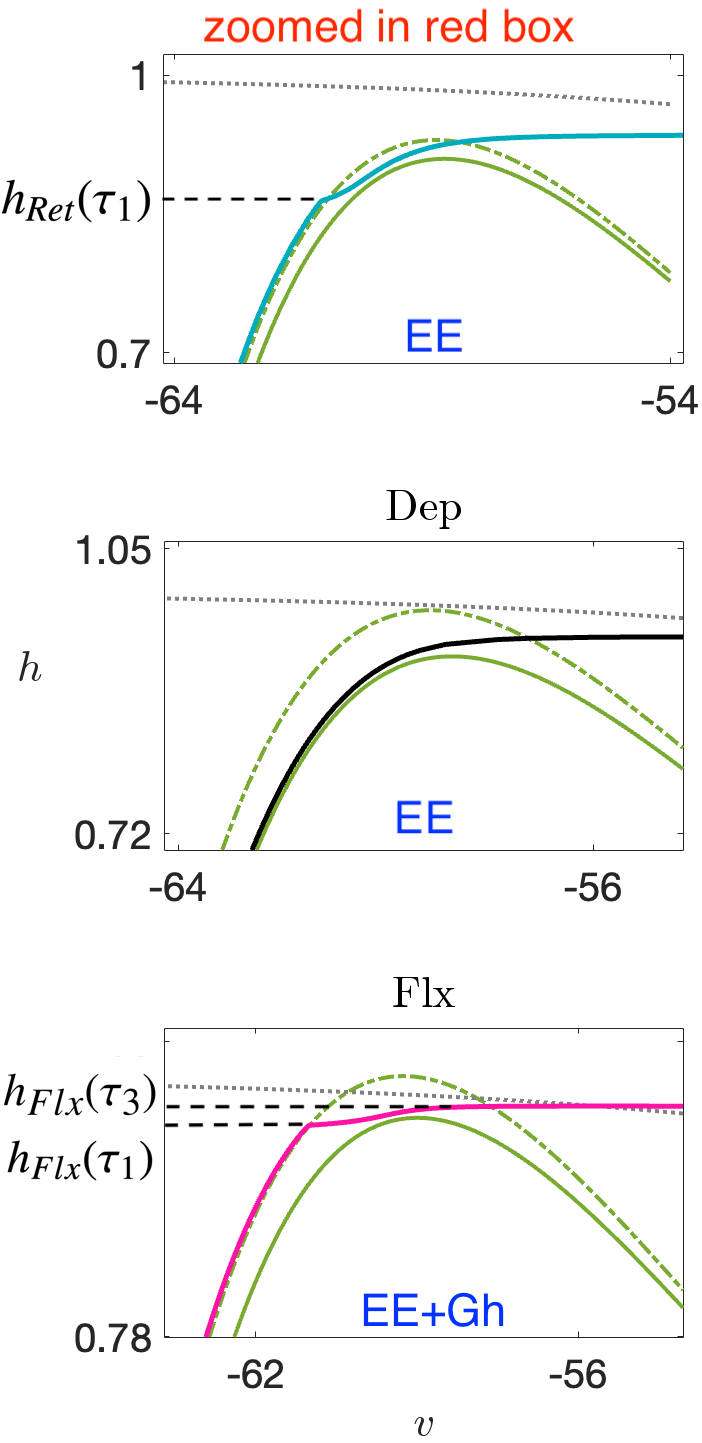}
\includegraphics[width=.5\textwidth]{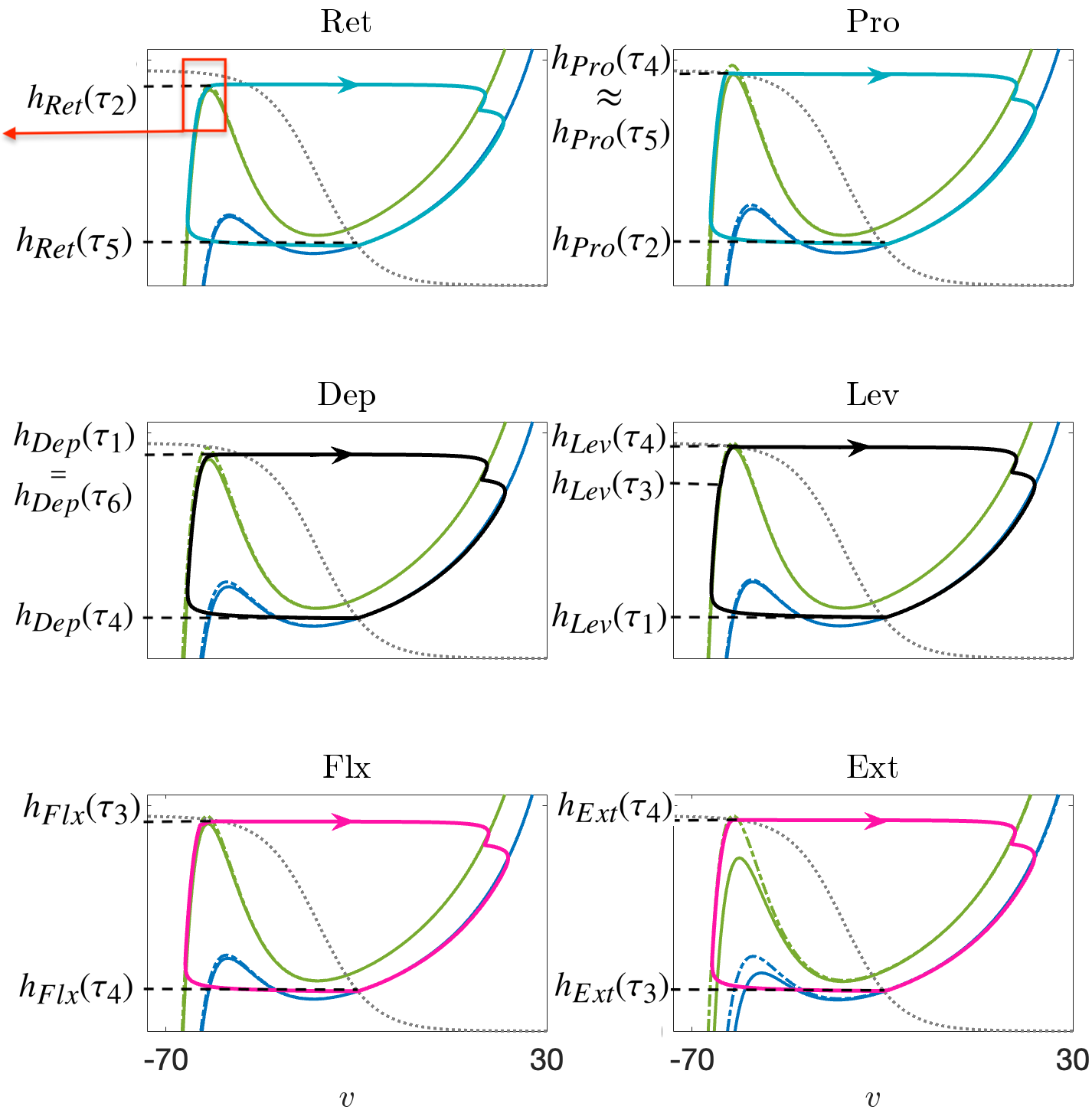}
\includegraphics[width=.23\textwidth]{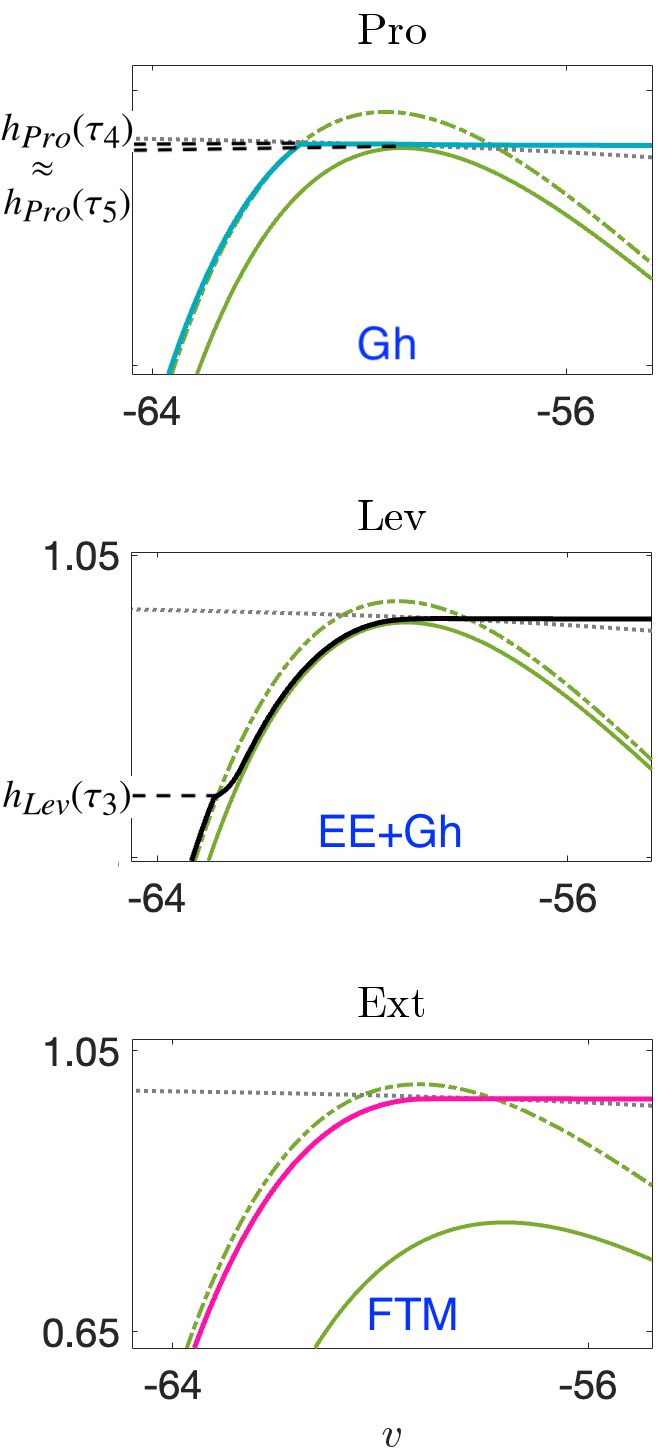}
    \caption{{\bf The $(v,h)$ phase planes for six coupled units, along with their corresponding zoomed versions.} Two pairs of cubic-shaped $v$-nullclines are depicted: the green pair of nullclines arises when antagonistic inhibition is on and the blue pair when it is off. In each same-color pair, the solid nullcline represents the scenario when excitatory inputs are on, and the dashed nullcline indicates when they are off. The dotted gray curve denotes the $h$-nullcline. Each phase plane includes the projection of a limit cycle, colored according to the scheme in Figure~\ref{fig:time-series-IN}. A zoomed view near the left knees of the inhibited nullclines is included for each phase plane to provide a more detailed visualization.  
    }
    \label{fig:both-6-phase-planes-IN} 
\end{figure}

Although the mesothoracic model  comprises a system of 12 coupled nonlinear ODEs, a helpful alternative to considering a 12-dimensional phase space is to visualize its trajectories in a set of six phase planes, one per CPG unit, each including the projection of the trajectory along with multiple sets of $v$-nullclines corresponding to different levels of inputs that affect the visualized unit at different times during a locomotor cycle (Section \ref{subsec:fast-slow}; Figure \ref{fig:both-6-phase-planes-IN}, with zoomed-in views shown in the outer columns). 
Specifically, each unit receives inhibitory input from its antagonistic partner within its joint core; the green pair of nullclines arise when inhibition is on and the blue (not visible in the zoomed-in plots) when it is off.
Moreover, each unit receives excitatory input from one or more sources; in each same-color pair of nullclines, the solid nullcline arises when excitation is on and the dashed when it is off (e.g., green pair in zoomed view in Figure \ref{fig:both-6-phase-planes-IN}). We will denote the attracting nullcline branches and associated structures for unit $i$, with $i \in \{ Pro, Ret, Lev, Dep, Ext, Flx \}$, by $v^i_L(h), v^i_R(h), (v_{CP}^i,h_{CP}^i)$, and so on; in doing so, we drop explicit reference to the input levels on which these values depend, although we will occasionally include these as needed.

Each zoomed nullcline visualization in Figure  \ref{fig:both-6-phase-planes-IN} focuses on the left knee, $(v_{LK}^i,h_{LK}^i)$, where the projected trajectory lies when the corresponding unit begins its transition from inactive to active, and hence reveals details of how each unit activates.  For each projection, we refer to unit $i$ as silent or inactive when its trajectory lies on or near the left branch of a 
$v$-nullcline, $\{ v = v_L^i(h;g^i_{exc},g^i_{inh}) \}$ for whatever inputs $(g^i_{exc},g^i_{inh})$ it receives, and as active when its trajectory lies on or near a right branch, $\{ v = v_R^i(h;g^i_{exc},g^i_{inh}) \}$.  
The six IN units' projections each have subtly unique features. 
One point of commonality is that we have tuned the model so that each unit's $v$- and $h$-nullclines intersect on the right branch of the $V$-nullcline, corresponding to a some level of tonic activation, in the absence of input (cf. blue dashed $v$-nullclines and dotted gray $h$-nullclines). 
When receiving reciprocal inhibition from its joint core partner, however, each unit has a left branch intersection and hence cannot activate on its own, with the exception of Ret and Dep.  Thus, if Pro (Lev) is active, then Ret (Dep) can nonetheless overcome the resulting inhibition and activate, suppressing its partner, but not vice versa.  This tuning reflects a default bias so that the leg is on the ground, not in midair, when at rest.  

\begin{figure}[h!]
    \centering
\includegraphics[width=1\textwidth]{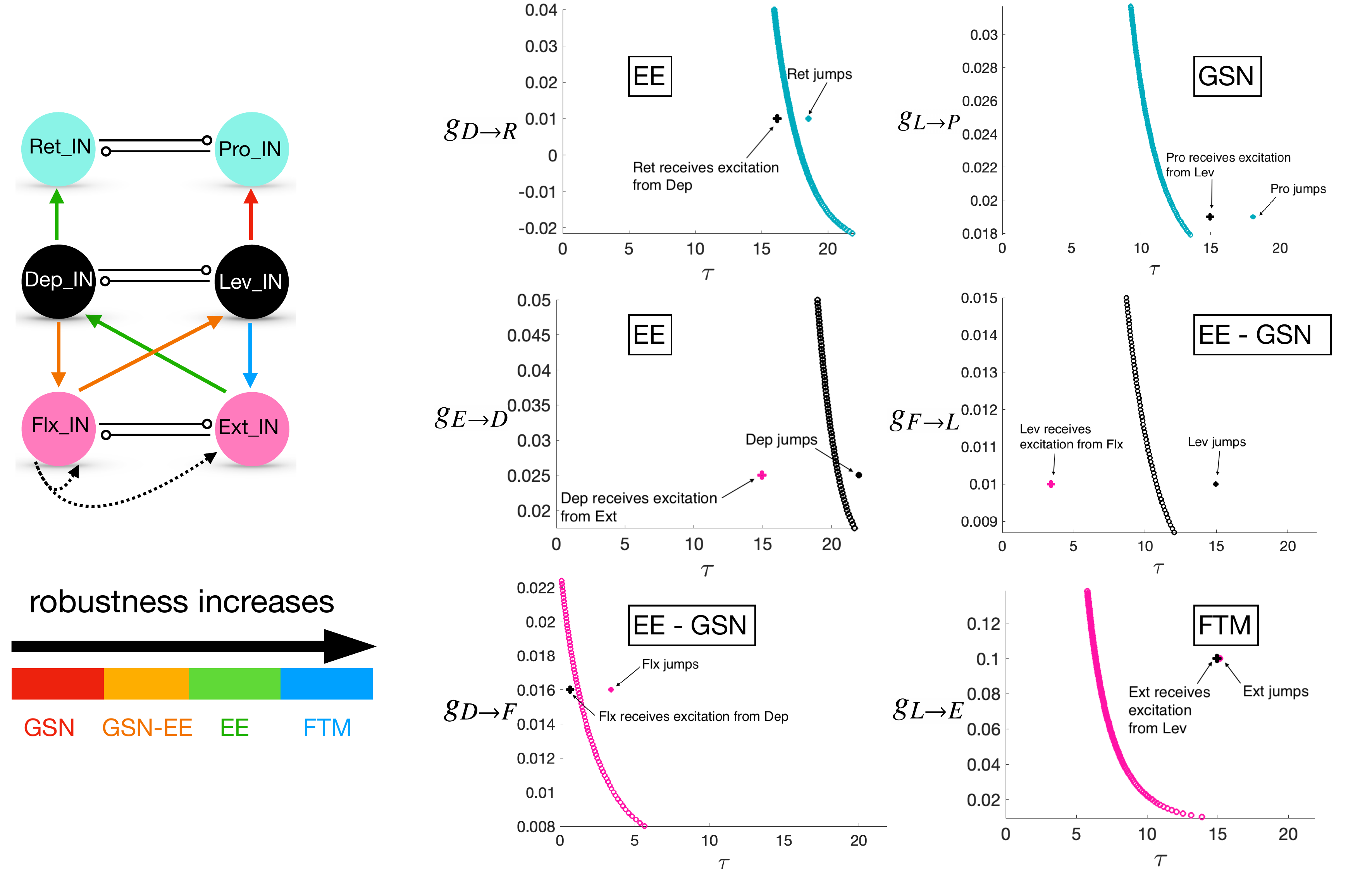}
    \caption{{\bf Determination of transition mechanisms and implications for parameter robustness.}  Left panel: Degree of robustness of parameters associated with excitatory connections depends  on the type of transition that they impact.  Right panel:  Determination of the transition type for the escape of each IN unit from the silent phase.  For each panel, the point indicated with a plus symbol indicates the excitation level that the unit receives and the time after its active phase entry when it arrives, while the dot is at the same excitation level but at the time after active phase entry when it jumps to the active phase.  The solid curves are the CTCs (see text). }
    \label{fig:Gh-FTM-EE} 
\end{figure}

 The zoomed views in Figure \ref{fig:both-6-phase-planes-IN} include labels indicating the type of transition involved in the escape of each unit in the network from its silent phase.
These labels can to some extent be inferred from inspection of the phase planes, but we performed a more careful, quantitative simulation experiment to solidify these determinations. 
Specifically, recall that when each unit first enters the silent phase, it travels along the $v$-nullcline $\{ v=v_L(h;\gsynin,0) \}$ corresponding to the inhibition that it receives from its joint core antagonistic counterpart.
For each, we computed the slow timescale time of passage $\tau$ along this nullcline from $h=h_{CP}(0,0)$, the approximate $h$-value at which it enters the silent phase, to $h=h_{LK}(\gsynin,g)$, for different values of $g$ (colored curves in Figure \ref{fig:Gh-FTM-EE}), starting with the $g$ value at which the saddle-node bifurcation that annihilates the unit's silent phase critical point occurs and increasing from there. We call the computed value the {\em clearance time} and the resulting curve the clearance time curve (CTC).  For our baseline rhythm, we then determined (a) the time $\tau_{in}$ after silent phase entry when each unit receives the excitatory input that eventually leads to its escape, (b) the strength $g_{ex}$ of this input, and (c) the time $\tau_{ju}$ after silent phase entry when each unit jumps up and hits the threshold $v=-30$.  
We marked the points $(\tau_{in},g_{ex})$ and $(\tau_{ju},g_{ex})$ in each plot with a plus symbol and a circle, respectively.

The locations of these points indicate the type of transition that each unit undergoes, based on the theory in Section \ref{sec:FTM-EE-Gh}.  For FTM, the $h$-value for the unit must be above $h_{LK}(\gsynin,g)$ when the excitation arrives, corresponding to the plus point $(\tau_{in},g_{ex})$ lying to the right of the CTC.  Moreover, the difference between $\tau_{in}$ and $\tau_{ju}$ must be small, corresponding to similar locations for the two marked points.  This case arises uniquely only for Ext.  
For Pro, the first condition holds, but certain additional features are present.  First, $\tau_{ju}-\tau_{in}$ is relatively large (as can be seen from the distance between the corresponding marked points).  Second, the $g$ value where input arrives (and where the marked points lie) is much closer to the minimal $g$ value where the saddle-node bifurcation occurs, such that the $v$- and $h$-nullclines are likely to be much closer together.  Thus, we recognize that this is a GSN transition. 
When $h < h_{LK}(\gsynin,g)$ at the arrival time of the excitation and hence $(\tau_{in},g_{ex})$ lies to the left of the CTC, we recognize that EE has occurred. In the cases of Ret and Dep, the jump-up point is close to the CTC.  For Lev and Flx, however, the distance in $\tau$ between the CTC and the jump-up point is comparable to the distance in $\tau_{ju}-\tau_{in}$ for the Gh case for Pro, and for Lev, the $g$ value itself is close to the saddle-node value.  Thus, we denote these two as EE-GSN transitions, given that they have elements of EE (excitation arrives before CTC is reached) and GSN (delay between crossing the CTC and jumping to the active phase).

We will now focus on the four phase transitions in the baseline rhythm: (1) activation of Ret and Flx, (2) activation of Lev and Ext, (3) activation of Pro, and (4) activation of Dep.   For convenience, will refer to the marked times $\{ \tau_1, \ldots, \tau_6 \}$ from Figure \ref{fig:time-series-IN} to represent the times when these transitions happen. That is,  we will start from the time $\tau_1$ when Dep activates and Pro, Ext are still active, with the subsequent transition times as follows: (1) $\tau_2$ and $\tau_3$, (2) $\tau_4$, (3) $\tau_5$, (4) $\tau_6$.

\medskip 

\noindent \underline{Transition (1)}: 
When Dep activates at time $\tau_1$, it excites both Ret and Flx.  The $h$-value for Ret  is not quite high enough to yield FTM, since $h_{Ret} < h_{LK}(\gexDR,\gsynPR)$, and hence the trajectory projected to the Ret phase plane undergoes an excursion on its solid green $v$-nullcline before Ret activates via an EE transition (e.g., at time $\tau_2$ in Figure \ref{fig:time-series-IN}; see Figure \ref{fig:Gh-FTM-EE}).
The $h$-value for Flx is approximately at the same level as the left knee of its solid green $v$-nullcline, $h_{Flx} \approx h_{LK}(\gexDF,\gsynEF)$, so any EE effect should be weak, yet the activation of Flx exhibits slightly more delay than does that of Ret (e.g., $\tau_3$ versus $\tau_2$ in Figure \ref{fig:time-series-IN}).  This delay arises via a  GSN effect, due to the proximity of the $v$- and $h$-nullclines for Flx 
(Figure \ref{fig:Gh-FTM-EE}).  Thus, the activation of Flx involves a blend of the EE and GSN mechanisms, which we can also think of simply as GSN transition preceded by a brief EE period.

\medskip 

\noindent \underline{Transition (2)}: Now, we consider the  state in which Ret, Dep, and Flx are active, corresponding to the main part of the stance phase, with their joint core partners suppressed (e.g, time interval  
$(\tau_3,\tau_4)$, Figure \ref{fig:time-series-IN}). 
In this situation, excitation signals go from Dep to Ret and Flx and from Flx to Lev and Ext.
Let us consider the nullclines for the three suppressed units: Pro, Lev, and Ext.  
Pro is inhibited by Ret and not excited, so its green dashed $v$-nullcline applies and there is a stable critical point, $(v^{Pro}_{CP}(0,\gsynRP),h^{Pro}_{CP}(0,\gsynRP))$, where that intersects its $h$-nullcline; thus, $(E1)_e$ fails for Pro and Pro cannot activate.
Although Lev is inhibited by Dep, the excitation from Flx causes its trajectory to jump from its dashed green to its solid green $v$-nullcline, 
which does not intersect its $h$-nullcline.  Thus, $(E1)_e$ holds for Flx and this represents an EE  configuration (Section \ref{sec:FTM-EE-Gh}).  After some time of transit up the left branch of this $v$-nullcline, at time $\tau_4$, Lev reaches $(v^{Lev}_{LK}(\gexFL,\gsynDL),h^{Lev}_{LK}(\gexFL,\gsynDL))$, experiences an additional GSN effect, and then fully activates, suppressing Dep, yielding another EE-GSN transition overall (Figure \ref{fig:Gh-FTM-EE}). 

Ext also receives excitation, along with inhibition, from Flx, so why does Ext not lead this transition?
The solid green nullcline for Ext in Figure \ref{fig:both-6-phase-planes-IN} corresponds to excitation from Lev to Ext, not from Flx to Ext, which is much weaker.  The excitation from Flx pushes the Ext trajectory just slightly below its purely inhibited (dashed green) $v$-nullcline.  Although it cannot be definitively seen in Figure \ref{fig:both-6-phase-planes-IN}, Ext still has a stable critical point, $(v^{Ext}_{CP}(\gexFE,\gsynFE),h^{Ext}_{CP}(\gexFE,\gsynFE))$, for this level of excitation, violating $(E1)_e$.  When Lev activates, it excites Ext, the Ext $v$-nullcline drops down to the solid green location,  with $h_{Ext} > h^{Ext}_{LK}(\gexFE+\gexLE,\gsynFE)$, 
and Ext immediately follows Lev into the active phase via FTM \cite{somers1993,rubin2002}
(Section \ref{sec:FTM-EE-Gh}; Figure \ref{fig:Gh-FTM-EE}). 

\medskip 

\noindent \underline{Transition (3)}: Once Lev and Ext activate, Lev excites Pro.  So why does Pro not activate together with these other units?  Notice in Figure \ref{fig:both-6-phase-planes-IN} that the projection of the trajectory to the Pro phase plane does jump away from the dashed green $v$-nullcline when this excitation arrives, at $t=\tau_4$.  Its solid green $v$-nullcline, corresponding to its being excited (by Lev) and inhibited (by Ret) does not intersect its $h$-nullcline; that is, the onset of excitation to Pro from Lev induces a saddle-node bifurcation of equilibria for Pro and $(E1)_e$ holds.  Nonetheless, the two nullclines lie extremely close together.  Their proximity yields a region in phase space where both $v$ and $h$ change slowly for the Pro unit, with $h_{Pro} \approx h_{LK}^{Pro}(\gexLP,\gsynRP)$.
The resulting GSN (Section \ref{sec:FTM-EE-Gh}; Figure \ref{fig:Gh-FTM-EE}) delays the onset of Pro activation, such as from time $\tau_4$ to time $\tau_5$ in Figure \ref{fig:time-series-IN}.

\medskip 

\noindent \underline{Transition (4)}: Just as the activation of Lev excites Pro, the activation of Ext excites Dep.  Dep does not have a GSN and $(E1)_e$ holds, so why does it not become active with Ext?  Recall that Dep first left the active phase in Transition (2), and the value of its $h$-coordinate is low when Ext activates (e.g., at time $\tau_4$ in Figure \ref{fig:time-series-IN}). Thus, even though the projection of the trajectory to the Dep phase plane lies on the solid,  excited $v$-nullcline of the Dep unit's green (inhibited) nullcline pair, which does not intersect its $h$-nullcline, it has $h_{Dep} \ll h_{LK}^{Dep}(\gexED,\gsynLD)$ and the projected trajectory must undergo a long excursion up the nullcline's left branch before reaching its knee.  When this excursion is completed, Dep activates (e.g., at times $\tau_1$ and $\tau_6$ in Figure \ref{fig:time-series-IN}), representing another EE transition (Figure \ref{fig:Gh-FTM-EE}).

\medskip

In summary, we see that the transitions in which units activate during this multi-phase baseline rhythm feature all three of the mechanisms that we presented previously (Section \ref{sec:FTM-EE-Gh}).  Of these, both EE and GSN mechanisms lead to a delay between the activation of one unit and the activation of another unit that it excites, while FTM does not.

The three mechanisms yield distinct predictions about sensitivity of activation timing to parameter variations (see also Section \ref{sec:FTM-EE-Gh}).  In the GSN mechanism, the delay time would be expected to be sensitive to small changes in the $v$-nullcline position, which could arise due to changes in excitation strength or in certain parameter values intrinsic to the unit's model equations.  Parameter changes that push the $v$-nullcline and specifically $h_{LK}$ even in a little bit lower would be expected to significantly shorten the delay in activation and could switch ghost transitions to FTM.

In EE, we expect more robustness to parameter variations as long as the  associated trajectory's jump to the excited $v$-nullcline's left branch happens far from its left knee, such that the slow excursion on the left branch is preserved as parameters are varied.  For excitation onsets relatively later within a unit's silent phase, however, a trajectory's $h$-value will be closer to that of the relevant left knee, such that parameter changes that affect nullcline position may switch a transition from EE, with $h < h_{LK}$ and some activation delay, to FTM, with $h > h_{LK}$ and a much shorter delay.  

Finally, FTM itself occurs abruptly on the fast timescale, so once a transition is within the FTM regime, activation timing should be robust to subsequent parameter changes unless they are so extreme that they switch the transition mechanism from FTM to one of the other types.
We will explore issues of robustness in much more detail in Section \ref{sec:robustness}.

\subsection{Robustness of stepping rhythms to model parameter values}\label{sec:robustness}

To find parameter sets that would yield the baseline rhythm, we tuned the three parameter classes $\{g_{ton,i}\}$, $\{\gsynin\}$, and $\{\gexji\}$ individually across units and coupled unit pairs.
Here, we consider the robustness of the rhythm to various forms of changes in the values of these key parameters.  We note that we are considering a model for a fictive locomotion rhythm produced in the absence of feedback signals and interactions across limbs that could promote robustness in a more biological setting; nonetheless, we assume that maintenance of a rhythm satisfying (P) across some range of parameter variations represents an important feature for our model.  We will consider four issues in the context of robustness:  (1) How robust is the baseline rhythm to changes in different classes of parameters, and why does robustness vary across these classes and specific elements within them? (2) Which parameters can be tuned to adjust rhythm frequency, and which rhythm features vary under this tuning? (3) When (P) is sensitive to changes in certain parameters, to what extent can variation of other parameters compensate to maintain (P) (cf. \cite{marder2015})?  (4)  Can inhibitory interactions across joint segments enhance robustness?   

\subsubsection{Varying parameters within classes}
\label{sec:robust_vary}
First, we tested the effect of multiplying all values in a selected parameter class by $(1+\epsilon)$ as we varied the scaling parameter $\epsilon$ over a small interval around 0.
We found that the most sensitive parameter class was $\{g_{ton,i}\}$, which could only tolerate scaling by $\mathcal{O}(10^{-3})$ values of $|\epsilon|$.

Although this uniform scaling was quick to implement as a preliminary step, and could represent biological variability in some factor that is shared across neurons or synapses, we would expect that variability in general could occur at the level of individual neurons and synapses.  Thus, we next turned to variation of each individual element of each of the three parameter classes. In our simulations, we tested for robustness with respect to variations of up to $\pm 5\%$ in parameter values.  We expected to find the least robustness with respect to parameters that affect the activation of Pro, since Pro activation relies most heavily on the ghost transition mechanism, which is sensitive to even small changes in $v$-nullcline positions.  Flx and Lev activations also feature a weaker form of the GSN effect, and hence seemed like another natural   source of parameter sensitivity. 

Overall, we found the most robustness with respect to variations in the excitatory, inter-segment $\{\gexji\}$ parameter class.  The network continued to produce a rhythm that satisfied (P) over $\pm 5\%$ variation of all such parameters except $\gexLP$, where robustness was limited, and $\gexFL$, which failed (P) specifically for $+5\%$ variation.  Since $\gexLP$ affects the position of the solid green Pro $v$-nullcline that is critical to Pro activation (\ref{fig:both-6-phase-planes-IN}),  non-robustness with respect to $\gexLP$ matches our expectations.  The much milder non-robustness with respect to $\gexFL$ is also consistent with our expectations based on the GSN mechanism of Lev. 

Robustness with respect to changes in $\{g_{ton,i}\}$ and the inhibitory, intra-joint  coupling $\{\gsynin\}$ was much more limited.
Within each of these classes, however, certain parameters yielded more robustness than others.  
The parameter $\gappp$ was the most sensitive of the external drive parameters, tolerating variations of less than $\pm 1\%$, again consistent with the sensitivity of  ghost transitions.  As this parameter was varied, the most fragile property of (P) was the synchronized inactivation of Pro and Ext.
Decreases in $\gappp$ delayed the activation of Pro and the inactivation of Ret.  This change propagated into delays in subsequent Ret activation, which  caused Pro to fall silent too long after Ext, and worsened across successive cycles.  Similarly, increases in $\gappp$ had the opposite effect, with successively earlier Pro activations and Ret activations, the latter of which desynchronized the inactivation of Pro and Ext.

We also found quite limited robustness in $\gappl$ and, to a slightly lesser extent, in $\gappf$.  Changes in $\gappf$ led to similar issues with desychronization of Pro and Ext inactivation, which we consistently found to be a vulnerable aspect of the rhythm, stemming from small changes in the timing of Pro activation through the GSN mechanism together with the lack of any direct connection to help coordinate Pro and Ext activity (see Figure~\ref{fig:CIRCUIT}). 
Decreases of $\gappl$ also harmed the relative timing of Pro and Ext inactivation: with small decreases, Pro inactivated too early relative to Ext, and with slighly larger drops it also inactivated too soon after Lev inactivation.  Lower $\gappl$ delays Lev's escape via its GSN, which correspondingly delays its recruitment of Ext by FTM, but Pro and Ret are less affected, which throws off the relative timing in the rhythm.  On the other hand, increases in $\gappl$ caused the rhythm to fall apart entirely.  These increases allowed for earlier Lev escape.  As a result, the excitation from Lev to Pro arrived when $h_{Pro}$ was smaller, making it more difficult for Pro to undergo its GSN escape.  This problem worsened across cycles, until eventually Pro failed to activate on one or more cycles.

The components of $\{\gsynin\}$ are mostly sensitive, with the exception of $\gsynLD$ and $\gsynFE$;
We expect that the stronger sensitivity to $\gsynin$ than $\gexji$ comes from the fact that the excitation level specifically affects the activation process.  Although certain units are excited throughout their active phases, voltage values lie near the excitatory reversal potential during these phases, so small changes in excitation levels do not have much effect there.  The inhibition to a unit from its joint partner, however, affects its evolution throughout its entire silent phase, potentially leading to stronger impact. 

\subsubsection{Compensation, neuromodulatory effects, and changes in period}
\label{sec:robust_compensate}
Although heterogeneity across individuals, neurons, and synapses could naturally occur at the individual parameter level, experiments have also revealed evidence of co-variation between values of different parameters \cite{marder2011,lamb2013,goaillard2021}. Moreover, there are biological mechanisms for which variations would translate into changes in multiple parameter values; for example, variation of the density of the synaptic neurotransmitter receptors on a post-synaptic neuron would affect the strengths of all inputs to that neuron that use that neurotransmitter, while variation of the rate of synaptic release by a pre-synaptic neuron would affect the levels of its coupling to all of its synaptic targets.
Such joint changes can occur biologically through the action of neuromodulators \cite{marder2001cpg,svensson2001,mackay2002}.
Therefore, we next considered the extent to which coordinated changes in the values of two or more parameters in the model could improve the robustness of its output pattern.

To start with, we considered the sensitive parameter $\gexLP$.
The model maintained a rhythm that satisfied (P) with nearly $50\%$ cuts in $\gexLP$ when these were accompanied by small, proportional increases in $\gappp$. These compensations maintained the ghost mechanism as needed for (P).
We could also maintain a (P) rhythm with up to $10\%$ increases in $\gexLP$ by matching each increase with a larger, proportional decrease in $\gappr$. With stronger $\gexLP$, Pro activates earlier and inactivates inappropriately before Ext.  A weaker $\gappr$ delays escape by Ret and hence leads to a prolonged period of Pro activation, which maintains the desired timing of Pro relative to Ext.  Despite the shorter delay from Lev activation to Pro activation in this regime, these changes preserve (P), as long as they are not too large.

In a similar vein, we found that we could achieve robustness with respect to at least $\pm 5\%$ variation of every one of the $g_{ton,i}$ parameters through compensating variation of another 1-2 parameters.
Variations in all the $g_{ton,i}$ parameters,   except for $\gappp$, could be fully compensated by either introducing stronger, same-sign variations in $\gsynin$ or by opposite changes in interjoint excitatory coupling $\gexji$. For instance,   
 variations in $\gappr$ could be fully compensated by corresponding, opposite changes in $\gexLP$, based on similar reasoning relating to the Pro active period, or by introducing stronger, same-sign variations in $\gsynPR$. 
Among the most sensitive parameters, 
compensation for $\gappp$ is achieved by increasing both $\gsynRP$ and $\gexLP$, while introducing stronger, same-sign variations in $\gsynDL$ together with same-sign variations in $\gexLP$ yielded 
robustness to $\pm 5\%$ variations in $\gappl$. 

Analogously, while almost all $\gsynin$ parameters are sensitive, they can tolerate greater variations if they are varied in coordination with $g_{ton,i}$ parameters. These results highlight the intrinsic duality between $\gsynin$ and $g_{ton,i}$: changes in one can be systematically offset by adjustments in the other, reinforcing their compensatory relationship. 

To complement this analysis, we specifically considered sets of parameter changes corresponding to groups of synapses with common sources or common targets, which could most naturally covary -- or be co-varied --  through biological mechanisms such as neuromodulation.  
For example, jointly scaling $\gexDR$ and $\gexDF$ by a common factor would correspond to modulating Dep outputs, and this manipulation maintained a rhythm satisfying (P).
We found a robustness to combined variations of strengths of excitatory outputs from most individual sources and to  combined variations of strengths of all excitatory inputs to most shared targets.  
Certain exceptions occurred that were, not surprisingly, associated with the Pro unit.  The parameter $\gexLP$ is sensitive, and in order to make rhythms robust to joint scaling of $\gexLP$ and $\gexLE$, the two excitatory outputs from Lev, we had to adjust either $\gappp$ or $\gappr$ as well.  As another example, the only excitatory input to Pro is $\gexLP$, which we have already seen is highly sensitive to variations.

\begin{figure}[hbt!] 
    \centering
\includegraphics[width=\textwidth]{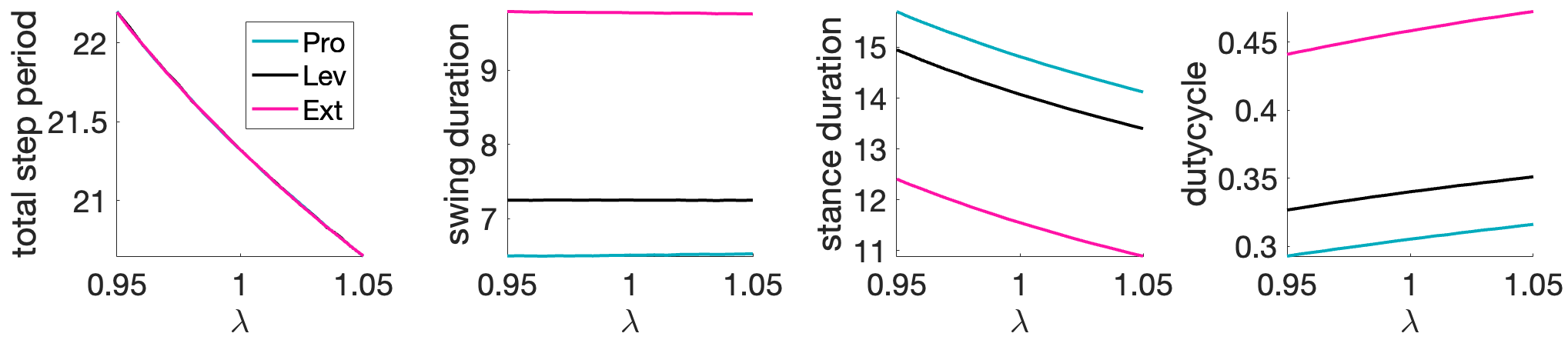}
\includegraphics[width=\textwidth]{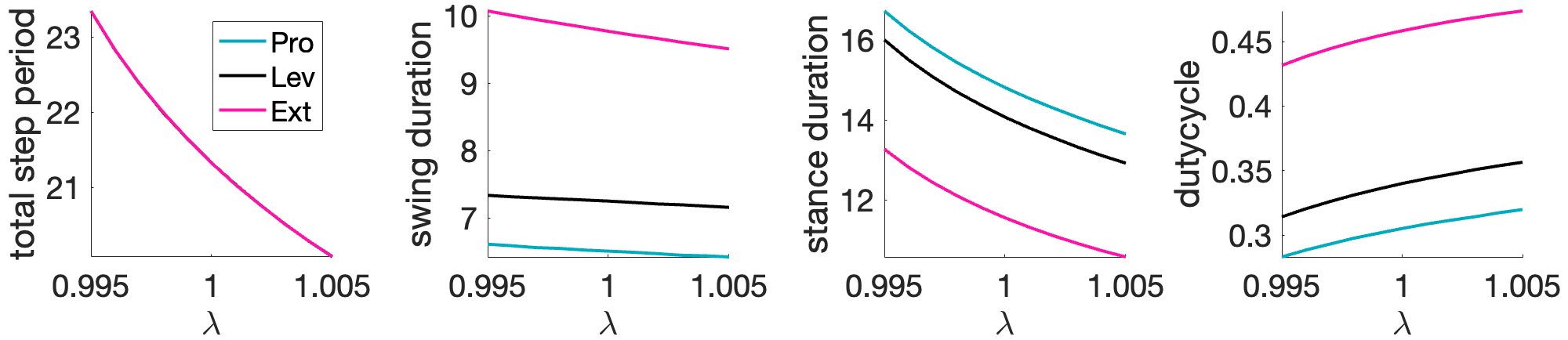}
    \caption{{\bf Impact of parameter variations on step cycle characteristics.} Period, swing, stance, and duty cycle of Pro, Lev, Ext are shown as the Flx excitatory output, namely $\gexFL, \gexFE, \gexFF$,  or Lev input, namely $\gappl$, $\gexFL$, $\gsynDL$,  change by a scaling factor $\lambda\in[0.95, 1.05]$ {(first row); and as $g_{ton,i}$ changes by a scaling factor $\lambda\in[0.995, 1.005]$ (second row).} 
    For each of Pro, Lev, and Ext, swing denotes the duration it is active in each cycle, while stance represents the duration that its respective antagonist -- Ret, Dep, or Flx -- is active in each cycle. Duty cycle represents the ratio of the active duration to the total period.}    
    \label{fig:varyFLX} 
\end{figure}

One interesting case that we highlight is joint scaling of all excitatory outputs from Flx, namely scaling $\gexFL, \gexFE, \gexFF$ by a common factor $\lambda$. We could achieve more than a 10$\%$ variation in the period of a rhythm that satisfies (P) using scaling factors of around $\pm 5\%$, i.e., $\lambda\in[0.95, 1.05]$.  
These changes in period arose through changes in the activation periods of Ret, Dep, and Flx (see Figure \ref{fig:varyFLX}), associated with the stance phase of the locomotor pattern, which is the aspect of the locomotor rhythm that varies under biologically observed period changes as well \cite{gabriel2007,vonuckermann2009}.  

Upon further investigation, we found that joint scaling of all the inputs to Lev, namely scaling $\gappl$, $\gexFL$, $\gsynDL$ by a common factor $\lambda\in[0.95, 1.05]$, achieves the same strong robustness of (P) and variation in period.  In fact, the key connection parameter underlying this effect as well as that from scaling excitatory Flx outputs is $\gexFL$.  
Note that when Flx activates, it excites Lev and Ext.  The next unit to escape after Flx is Lev, and this escape occurs through the EE mechanism associated with the excitation from Flx, with the actual moment of escape happening when the trajectory reaches $h_{Lev} = h_{LK}(\gexFL,\gsynDL)$. The $h$-coordinate of this knee is modulated by $\gexFL$; stronger excitation leads to a lower $h_{LK}$ and hence earlier escape, which corresponds to a shorter silent phase of Lev and active phase of Dep (Figure \ref{fig:varyFLX}, stance panel) and hence contributes to the shortening of period. Once Lev escapes, it excites Ext, which escapes via FTM, a highly robust mechanism, leading to a shorter active phase of Flx, and it also excites Pro, which shortens the Ret active phase, both of which further decrease the period.   On the other hand, the fact that Lev, Ext and Pro jump up with different timing does not affect the $h$ values where their respective joint block partners, Dep, Flx, and Ret, return to the silent phase, since in all cases, that happens very close to a critical point on the right branch of the corresponding $v$-nullcline, associated with condition (E3). Thus, Dep, Flx, and Ret will not undergo changes in their silent phase durations before their next escape, and correspondingly there are no changes in the Lev, Ext, and Pro active phase durations (Figure \ref{fig:varyFLX}, swing panel).  In fact, there is no other parameter that we have identified that can induce such a robust variation of cycle period, because of the unique role of the excitation from Flx to Lev.  Indeed, although Dep and Ret both activate via an early excitation mechanism involving excitation from Ext and Dep, respectively, and the existence of the rhythm is robust to changes in $\gexED$ and $\gexDR$, variation of these parameters cannot achieve significant change in period because they lead to compensatory effects that cascade through the active phase durations of all units in the network.

Finally, from a conceptual point of view, we would expect that top-down signaling should be able to modulate the period of locomotor output, so that changes in sensory inputs processed in other brain areas can lead to appropriately altered stepping.  In agreement with this expectation, we find that scaling {all $g_{ton,i}$}
 between 0.995 and 1.005 gives a variation in period from about 
 {20.08} to about 
{$23.35$} time units, with changes mostly in stance duration (Figure~\ref{fig:varyFLX}(second row)).  Thus, while $g_{ton,i}$ can only tolerate parameter changes that are small in magnitude, this model can nonetheless capture top-down frequency control.

\subsection{Interjoint inhibition}
\label{subsec:extension_inhibition}

Following the model developed in \cite{daun2011}, the interjoint coupling considered to this point in our analysis has been excitatory, with no interjoint inhibitory coupling included. 
The experimentally derived interaction scheme for CPG neurons in the mesothoracic leg of stick insects does not rule out the possibility of direct, interjoint inhibitory connections in addition to the excitatory ones, however.  For example, an activating influence of Dep on Ret could be implemented entirely by excitation or could involve an element of inhibition from Dep to Pro, which helps to relieve Ret from Pro's inhibitory influence.
To consider the impact of such connections on the generation and maintenance of robust, idealized rhythms, we next introduced additional inhibitory connections into our model, which we denoted by $\ginji$, with one new connection complementing each of the excitatory connections in the baseline model.  Note that all of these new connections comprise interjoint inhibition except that we added a $\ginFE$ component as well.
For simplicity, the $\ginji$ values for all new inhibitory connections were set as a fixed fraction of the excitatory connections inducing the same effective influence (e.g., $\ginDP$ was this fixed fraction of $\gexDR$).

Incorporating these connections preserved the rhythm (P) depicted in Figure~\ref{fig:time-series-IN}. Additionally, the sensitivity of most previously-included parameters remained unchanged, except for two: $\gsynFE$ and $\gappe$, both of which impact the ability of Ext to activate and exhibited  noticeably improved robustness to perturbations (i.e., by at least $\pm 5\%$).
This enhanced robustness was achievable only with the inclusion of the $\ginFE$. Specifically, 
decreases in $\gsynFE$ had caused (P) to be lost, and adding $\ginFE$ compensated for decreases in $\gsynFE$. Also, increases in $\gappe$ had eliminated (P), and the addition of inhibitory input $\ginFE$ compensated for the effectively excitatory effects of increases in $\gappe$. 
Similarly, since $\ginDE$ contributed to the level of inhibition that Ext encountered when trying to escape, its introduction improved the robustness of $\gappe$ as well.

\begin{figure}[hbt!] 
    \centering
\includegraphics[width=.9\textwidth]{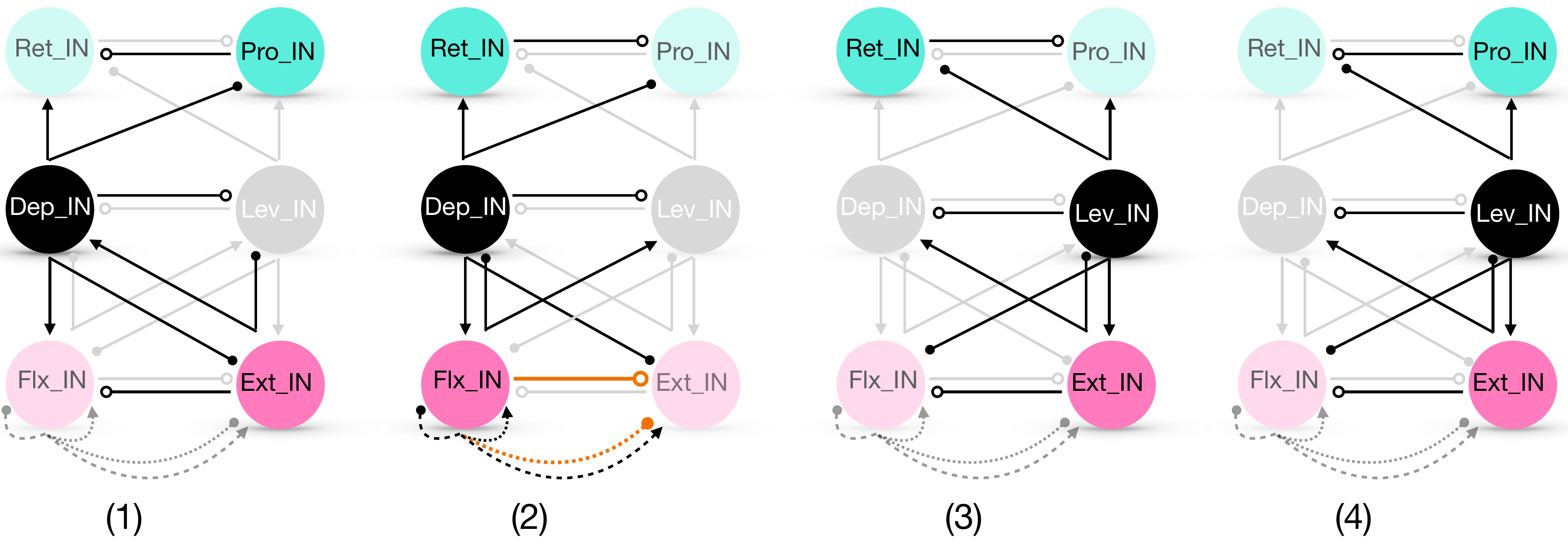}
\caption{{\bf The CPG circuit with inter-joint inhibition across four successive phases of a single step cycle.} The schematics shown are as in Figure~\ref{fig:time-series-IN}, with bold or pale coloring to indicate the nodes that are active or inactive in each phase and the corresponding outgoing connections. The inhibition from Flx to Ret in the second phase (indicated in orange) is the only coupling that improves the robustness of $\gsynFE$ and $\gappe$. See the text for more details.}
 \label{fig:inter-joint-inhibition} 
\end{figure}

None of the other additional inhibitory connections yielded enhanced robustness. The explanation for this lack of effect becomes apparent from inspection of Figure \ref{fig:inter-joint-inhibition}.  As an example, consider the inhibitory connection from Dep to Pro, with strength $\ginDP$.  This inhibition is on when Dep is active, in phases (1) and (2) in Figure \ref{fig:inter-joint-inhibition}.  Yet, when it becomes time for Pro to escape, in phase (3), Dep is inactive, this inhibition is no longer present, and hence overall this inhibition does not have an impact on robustness of parameters that influence the activation of Pro.  The timing of when all of the other added inhibitory connections are active is analogous; that is, each of these is off at the time when its target needs to escape, resulting in a minimal impact on robustness of other parameters.

\subsection{Extending the model to include motoneurons}\label{subsec:extension_MN}

In the original model of the mesothoracic limb network developed in \cite{daun2011} and illustrated in Figure~\ref{fig:CIRCUIT} (middle panel), each component consists of a pool of INs and a pool of MNs. In this network, all inhibitory synapses originate from INs, while all excitatory synapses originate from MNs.
In this work, we removed the MNs and modeled six pools of interneurons using a system of 12-dimensional ODEs. The primary motivation for excluding the MNs was to decrease the number of free parameters to tune and to   reduce the model’s sensitivity to parameter variations.
Despite this modification, we 
allow INs to send out both excitatory signals, with a reversal $\Eexc$, and inhibitory signals, with a distinct reversal $\Einh$.

\begin{figure}[h!]
    \centering
\includegraphics[width=\textwidth]{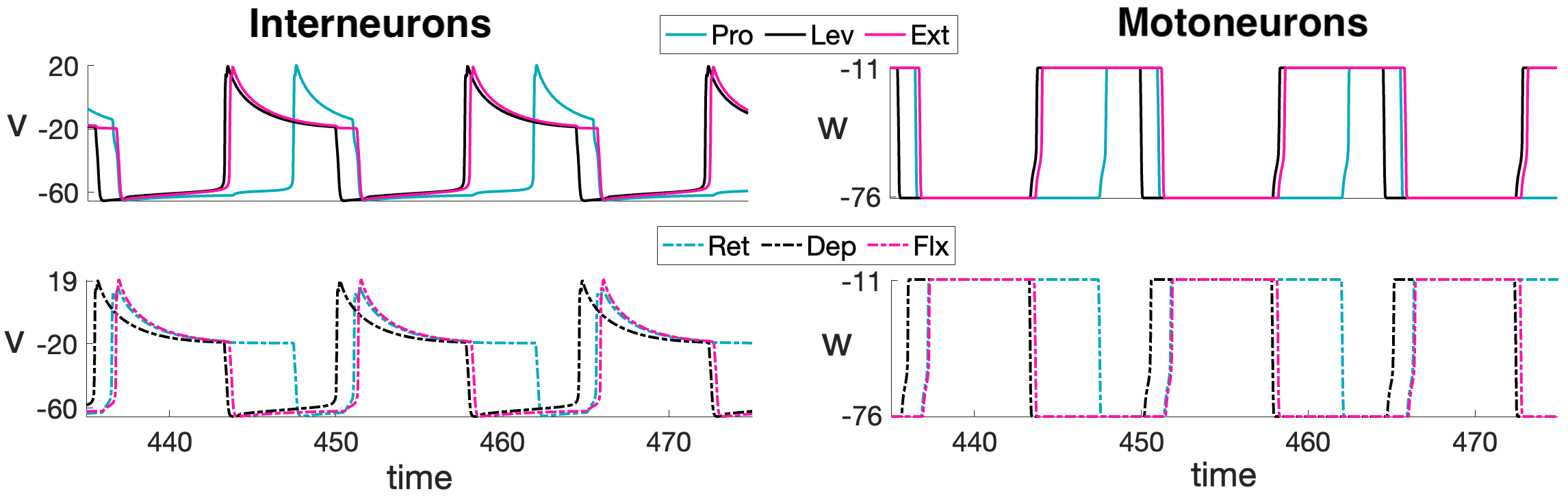}
    \caption{{\bf Voltage time courses of INs (left) and MNs (right) during a baseline locomotor rhythm with timing properties (P) consistent with those observed in stick insect fictive locomotion.} Top plots: activity of synergistic INs and MNs.  Bottom plots: activity of the respective antagonists, depicted in matching colors. The underlying circuit corresponds to the diagram shown in Figure~\ref{fig:CIRCUIT} (middle panel), and the parameters are specified in the columns for  ``IN-MN, Escape" of Tables~\ref{table:params_inh_exc_coupling}-\ref{table:params_exc_coupling} in Section~\ref{Sec:appendix}.}
\label{fig:Time_series_IN_MN} 
\end{figure}

In the extended model that includes MNs, all interjoint excitatory inputs 
originate from MNs instead of directly from INs (Section \ref{sec:circuit-CPG-MN}).  
MNs themselves activate when their joint block IN partners do, because activation of the IN partner inhibits the other IN in the joint block, thus disinhibiting the MN.
Figure~\ref{fig:Time_series_IN_MN} presents the time series of INs and MNs for an ideal rhythm (P) generated using the baseline parameters provided in Tables~\ref{table:params}-\ref{table:params_exc_coupling} in Section~\ref{Sec:appendix}. Although an ideal rhythm (P) can be achieved by this parameter set,  it is less robust than the ideal rhythms generated with the IN-to-IN excitatory connections. The reason for this difference lies in the behavior of the MN voltages, which exhibit only two states -- constantly active ($\sim$ -11) or constantly inactive ($\sim$ -76) --  resembling a step function.  In contrast, IN voltages display gradual transitions between active ($\sim$  +20 to -20) and inactive ($\sim$ -65 to -56) states. The abrupt switching in the MN model arises because, following past work \cite{daun2011} and based on the fact that MNs lack intrinsic rhythmicity \cite{mantziaris2020}, it has only one-dimensional dynamics that evolves on the fast timescale, with MN voltage rapidly equilibrating to steady state when the input to an MN turns on or off.  Moreover, each MN receives input from only one source.  Together, these features eliminate the possibility of EE effects and reduce the model’s tolerance to parameter variations, making it more sensitive to changes in input parameters.

\section{Generation of stepping rhythms in insects' limbs with transitions by release}\label{sec:release}

\subsection{A baseline model}

In a release transition, an active neuron that had been inhibiting its joint partner ceases to provide this inhibition, usually because it leaves the active phase.  This loss of inhibition releases its partner to become active (Section \ref{subsec:escape_release}).  In the escape case without interjoint inhibition, a critical point can be present on the the left branch $\{ v=v_L(h) \}$ when a neuron is inhibited, which prevents it from escaping prematurely. As we have seen, an excitatory input can then recruit that neuron through fast threshold modulation, a ghost transition, or an early excitation transition.
In release, we also expect that transitions will relate to changes in input, since synaptic interactions establish phase relationships between units. In the release case without interjoint inhibition, the change of input  that induces release must be a loss of excitation to an active neuron that triggers it to become inactive, rather than a gain of excitation by an inactive neuron.  The loss of excitation may yield an immediate fast threshold modulation, but as we have seen, some phase transitions in the locomotor rhythm must involve delays relative to a change in excitation. This requirement is problematic because the voltages of active neurons are near the reversal potential of excitatory synapses.  Hence, changes in excitation have relatively little effect on nullcline position, which limits the range of phase differences that can occur.

A specific difficulty in this case is associated with the delay between Dep activation, due to release by Lev, and activation of Flx and Ret, due to release by Ext and Pro, respectively.  That is, Ext and Pro need to remain active for roughly the same extended period after $\gexLE$ and $\gexLP$ go to 0,  yet there is no mechanism to coordinate when they jump down.   Thus, we instead turn to the case of release where interjoint inhibition is included.

\begin{figure}[h!]
    \centering
\includegraphics[width=0.8\textwidth]{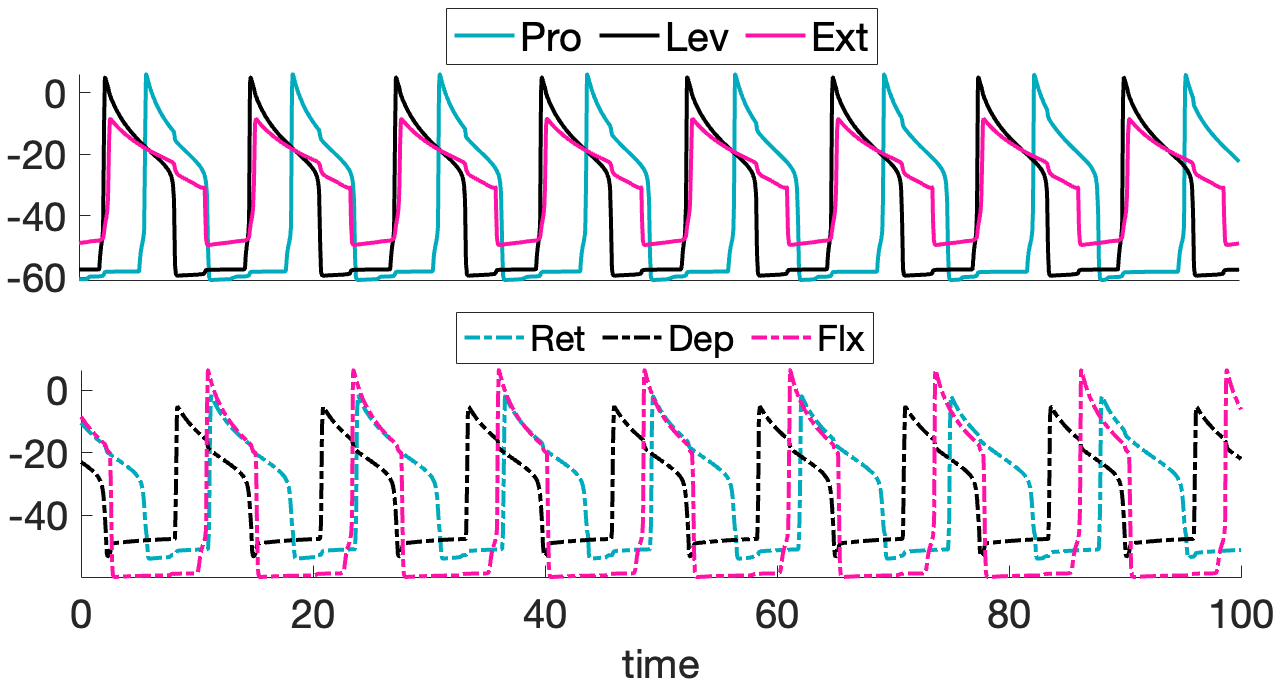}
    \caption{{\bf Voltage time courses of INs during a baseline locomotor rhythm with timing properties (P) consistent with those observed in stick insect fictive locomotion, from a model with interjoint inhibition.} The top plot shows the activity of synergistic INs, while the bottom plot displays the activity of their respective antagonists, depicted in matching colors. The underlying circuit corresponds to the diagram shown in Figure~\ref{fig:CIRCUIT} (right panel), and the parameters are specified in the columns  for  ``only IN, Release" of  Tables~\ref{table:params_inh_exc_coupling}-\ref{table:params_inh_coupling} in Section~\ref{Sec:appendix}.}
\label{fig:time_series_release} 
\end{figure}

\subsection{A model with interjoint inhibition}
\label{sec:releaseIN}
\begin{figure}[h!]
    \centering
    \includegraphics[width=0.45\textwidth]{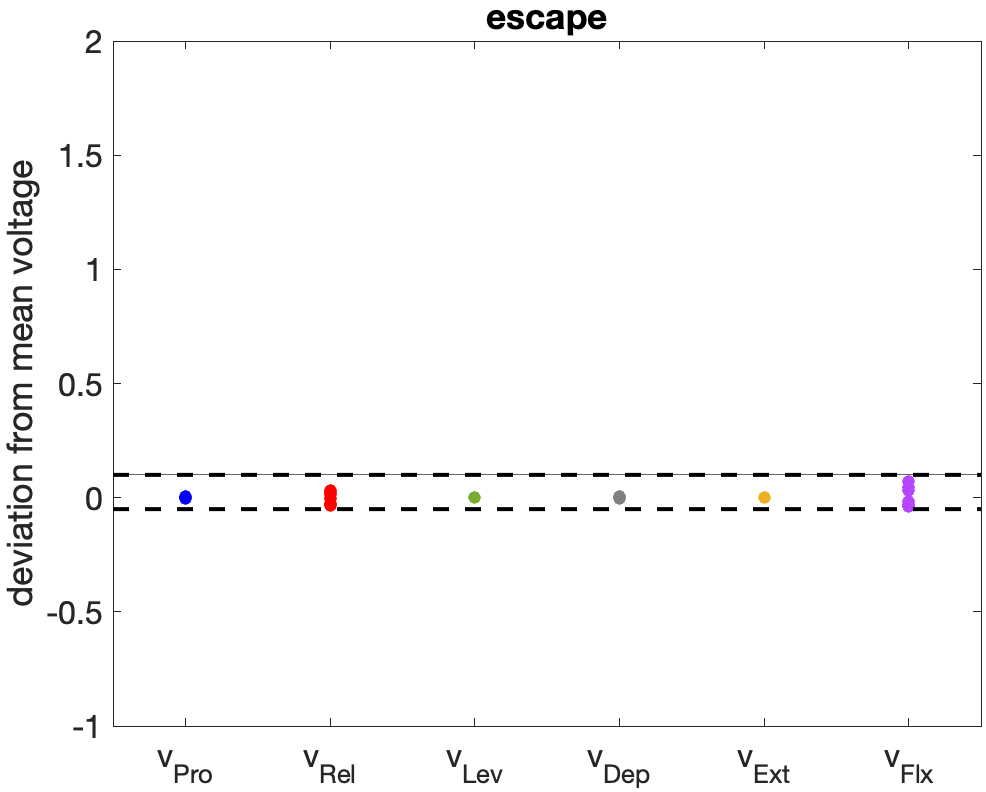}
    \includegraphics[width=0.45\textwidth]{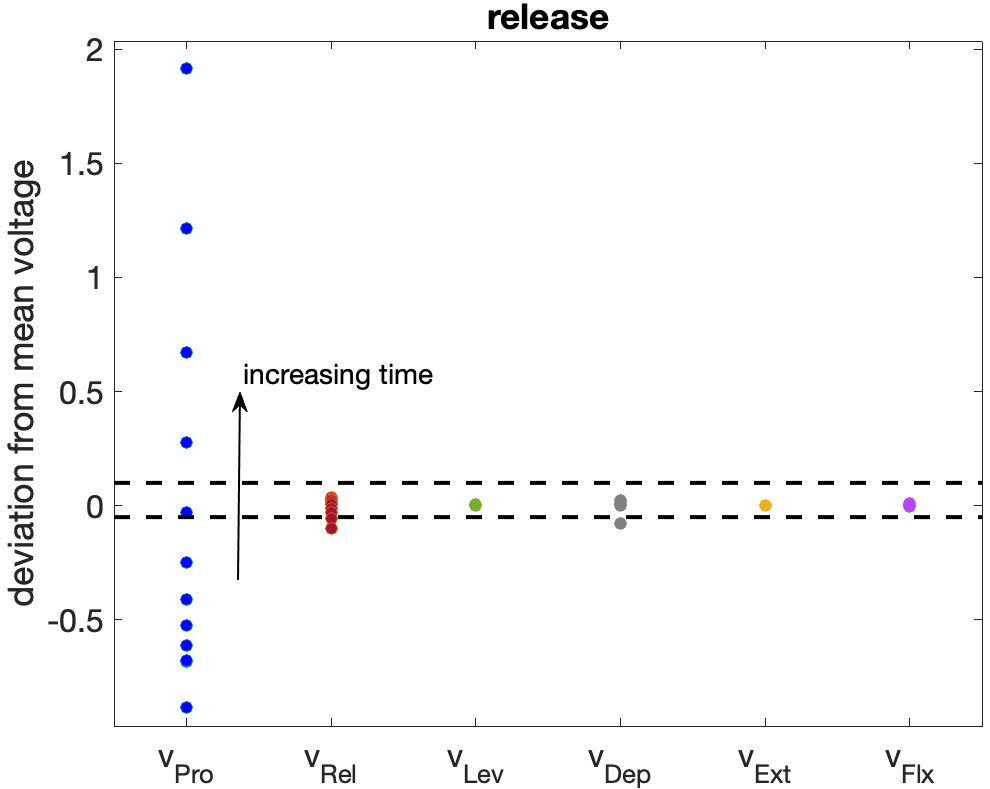}
    \caption{{\bf Deviations of voltage variables from the mean values computed from Poincar\'{e} section crossings, defined by $v_{Ext}=-30$ with $\dot{v}_{Ext}<0$, over 12 successive oscillation cycles.}  The non-zero variability in the escape case arises from numerical errors for a  periodic solution.  The more substantial variability in the release case, especially the increase in $v_{Pro}$ across successive cycles, arises from aperiodicty of what is actually only a metastable periodic solution.  }
    \label{fig:combinePoincare} 
\end{figure}

Figure \ref{fig:time_series_release} shows a locomotor rhythm with release transitions in a circuit with interjoint inhibition.  
Unlike the case of escape transitions, this rhythm is not actually stable but rather metastable, lasting for over 100 time units before failing to meet the criteria (P) that we seek.  This metastability is illustrated in Figure \ref{fig:combinePoincare}.  To generate this figure, we set up a Poincar\'{e} section at $v_{Ext}=-30$ and captured the $v$ values of all neurons in the network when the section was crossed with $\dot{v}_{Ext} < 0$.  We did this over 12 oscillation cycles for our baseline parameter sets, averaged the values of each $v$ variable, and then plotted the 12 deviations from the mean.  We can see that in the escape case, there is almost no variability, although the fact that $v_{Ext}=-30$ is not exactly achieved in any numerical simulation introduces a small amount of jitter in the calculated values  
In the release case, however, there is a clear and substantial increase in $v_{Pro}$, relative to its mean, over the successive cycles, due to a lack of true periodicity of the pattern.   
We were not, however, able to find parameters that provided a substantially longer stable persistence with release transitions.

\begin{figure}[h!]
    \centering
\includegraphics[width=0.8\textwidth]{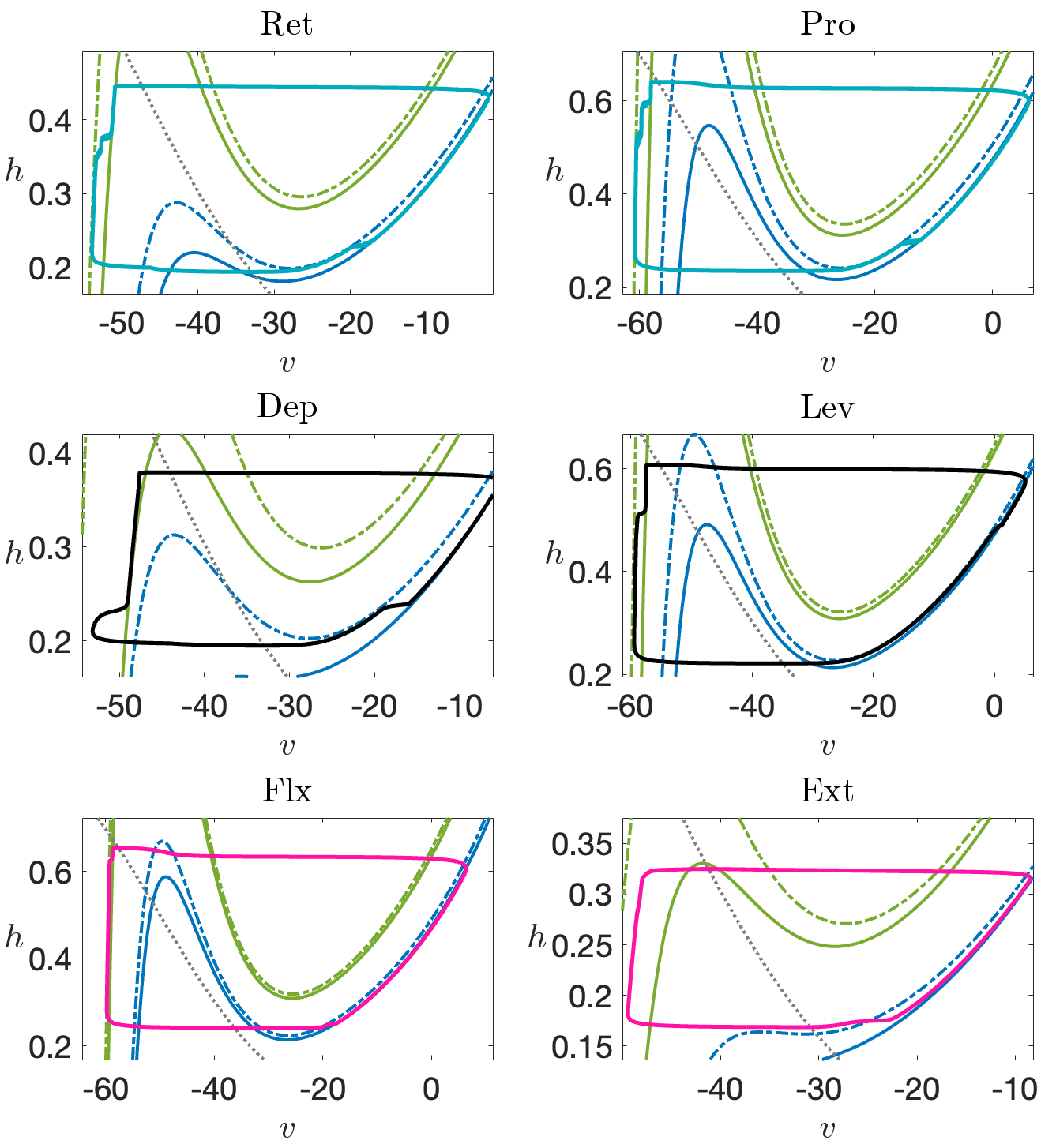}
    \caption{
    {\bf The $(v, h)$ phase planes for six coupled units in the release case.}
    The nullclines and trajectories follow the same conventions described in Figure~\ref{fig:both-6-phase-planes-IN}.
    The underlying circuit corresponds to the diagram shown in Figure~\ref{fig:CIRCUIT} (right panel), with parameters specified in the columns  for ``only IN, Release" of  Tables~\ref{table:params_inh_exc_coupling}-\ref{table:params_inh_coupling} in Section~\ref{Sec:appendix}.  For additional discussion, see Section \ref{sec:releaseIN}.
    }
\label{fig:6PP_release} 
\end{figure}

With inhibitory connections between joints, inhibitory inputs from a unit in another joint can induce a neuron to leave the active phase, thereby releasing its joint partner to activate.  The nullcline projections of the resulting release solution are complicated, since each neuron can receive various combinations of excitation and inhibition during each oscillation cycle  (Figure \ref{fig:6PP_release}).  In the basic oscillation pattern, Dep releases Lev due to a loss of excitation from Ext and a gain of inhibition from Flx.  The resulting loss of excitation from Dep to Flx, along with the gain of inhibition from Lev to Flx, causes Flx to jump down as well through an active phase FTM effect (Figure \ref{fig:6PP_release}, lower left).  At this stage, Lev, Ext, and Ret are active (e.g., around time 30 in Figure \ref{fig:time_series_release}).  The next event should be the jump down of Ret, about halfway through the Lev active phase.  The loss of excitation from Dep to Ret and the gain of inhibition from Lev to Ret should cause Ret to jump down, and this is delayed relative to the jump-down of Flx because the trajectory projected to $(v_{Ret},h_{Ret})$ ends up on the right branch of the appropriate $v_{Ret}$-nullcline and has to travel down to the right knee before its active phase terminates. Once this happens, Lev, Ext and Pro are active, and the next release should be instigated by Lev.
There are a loss of excitation from Flx to Lev and a gain of inhibition from Ext to Lev when Flx jumps down and Ext activates, but Lev must stay up long beyond that moment, again based on traveling down a $v$-nullcline right branch (Figure \ref{fig:6PP_release}, middle right).  
Once Lev jumps down, this removes excitation from Pro and Ext and introduces inhibition from Dep to these targets, but for 
a successful rhythm that meets our benchmarks, neither of them can jump down right away.  
As noted above, achieving a near-simultaneous jump-down of Pro and Ext, despite the absence of active phase critical points or other such structures that might promote phase locking, becomes a natural source of non-robustness in the release-based rhythm, with small changes in the connection strengths involved leading to loss of the desired rhythm structure.

In summary, because the inhibitory reversal potential lies far from the voltages that arise in the active phase, tuning of interjoint inhibition provides a much greater flexibility in controlling nullcline positions in the active phase than does tuning of interjoint excitation.
Yet, we have seen that this flexibility comes at a cost.  For example, when Lev jumps down and Dep activates (Figure \ref{fig:6PP_release}, middle row), not only does Ext lose excitation from Lev, but it now becomes inhibited by Dep, which makes it even more difficult to delay the jump-down of Ext (Figure \ref{fig:6PP_release}, lower right). On the other hand, if Ext falls down before Pro, then Flx is released and becomes active, which inhibits Dep.  This causes a small decrease in the inhibitory connection strength from Dep to Pro, $\gsynDP$, which makes it harder for Pro to release Ret and hence produces a further delay between the jump-down of Ext and that of Pro.
This latter perturbation will grow over successive cycles; indeed, the fact that Pro and Ret do not send excitation or inhibition to units in other joints simplifies the tracking of effects of perturbations to these units.  Specifically, if Pro and Ret respectively enter their silent and active phases later, then Ret has less time in the active phase with excitation from Dep, such that it evolves more slowly there and ends up releasing Pro later, meaning that Pro has less time in the active phase with excitation from Lev and jumps down to its next silent phase even later: there is an overall growth of the perturbation across cycles.
Since the release rhythm is metastable rather than stable and we identified several sources of sensitivity among the connection strengths associated with this rhythm, we did not perform further robustness studies for the release case.

%~~~~~~~~~~~~~~~~~~~~~~~~~~~~~~~~
\section{Discussion} \label{sec:discussion}
In this work, we set out to 
explore the capability of a minimal model of non-spiking neural units \cite{ludwar2005} to produce an idealized rhythmic output pattern associated with the stepping pattern of the stick insect mesothoracic leg \cite{buschges1994,fischer2001,akay2004,daun2009,bidaye2018}.  The unit dynamics in the model derived from a previous study \cite{daun2011}, with updates in parameter tuning.  The connectivity among units was based on the existence of an antagonistic muscle pair and associated CPG unit (joint core) for each joint \cite{buschges2008,mantziaris2020}, with interactions among these units determined experimentally \cite{bidaye2018}.
Our results show that these elements suffice to produce a sustained, stable stepping rhythm, as long as model parameters are tuned to yield transitions in which component of a joint core is active that are based on escape \cite{wang1992,skinner1994moa}, rather than release.
This rhythm is robust to certain parameter changes but not others, and its period can be tuned, via variation in the stance phase duration \cite{gabriel2007,vonuckermann2009}, through parameter variations that may be induced biologically by neuromodulators.  Thus, our model suggests new ideas about stick insect locomotor rhythm generation, generates predictions for future experimental testing,  and can serve as a building block for future computational studies.  

While an idealized rhythm matching our property (P) was produced in a previous model study upon which ours builds \cite{daun2011}, that rhythm only persisted transiently.  One of the challenges that we overcame to achieve a stable form of this rhythm was the fact that the target pattern features various phase delays between the arrival of excitatory input to a unit and the activation of that unit.  We go beyond past analyses of escape to distinguish three types of escape events -- FTM \cite{somers1993}, EE, and ghost \cite{izhikevich,hastings2018} -- that represent distinct points on a continuum of possible transitions and discuss how they can yield distinct phase delays. 
We show that our model's rhythm includes all of these types and analyze both how they relate to the phase delays in its rhythmic output and how they impact the robustness of the desired rhythm properties (P) to parameter variations.  For example, because Ext activates by FTM, the rhythm is relatively robust to changes in parameters associated with Ext excitability and inputs to Ext.   On the other hand, because Pro activates  through the ghost mechanism, which is sensitive to the distance between its voltage nullclines with and without excitatory input, the rhythm lacks robustness to variation of parameters related to inputs to Pro; $g_{ton,Lev}$ and $g_{ton,Flx}$ are similarly sensitive parameters.

Interestingly, we found that for those parameters for which small changes led to the loss of property (P), compensatory changes in other parameters within the classes $\{ \gsynin \}_{i,j}$, $\{ \gexji \}_{i,j}$, and $\{ g_{ton,i} \}_i$
could restore (P).  This result is not trivial because these different parameter classes have distinct properties.  That is, unlike the others, each $\gsynin$ has a negative reversal potential and hence becomes smaller (larger) in magnitude when the postsynaptic neuron is in the silent (active) phase.  Moreover, the $\{ g_{ton,i} \}_i$ impact unit dynamics all of the time, whereas elements of the other two parameter classes are only relevant when the presynaptic neuron is active. Our findings of compensation extend previous computational work showing that synaptic conductances and ion channel conductances can play compensatory roles \cite{grashow2010}.  while experiments have found evidence for compensatory variations in ion channel conductances within individual neurons \cite{maclean2005,marder2006,marder2011},  compensation across synaptic inputs of different types or between intra-circuit and top-down input strengths has not, to our knowledge, received such attention.  Nonetheless, various factors including neuromodulation and temperature could modulate input resistance and hence cause synaptic conductances to covary, so our results are consistent with the prediction that stick insect locomotor CPG dynamics should exhibit robustness to temperature variations, as observed experimentally in other CPGs \cite{yamaguchi2008,tang2012}.

Neuromodulators can also have more targeted impacts on the signaling by or to neuron populations \cite{marder2001cpg,svensson2001,mackay2002}.  Simulating such an effect, we found that a common scaling either of the outputs from Flx or of the inputs to Lev are uniquely effective, among variations in synaptic parameters, at modulating the period of the rhythm without compromising property (P), and that the changes in period occur through variation of the stance phase in which Ret, Dep, and Flx are active (interval $(\tau_3,\tau_4)$ in Figure \ref{fig:time-series-IN}).  We  explained these properties based on the interactions and phase transitions throughout the model circuit (Section \ref{sec:robust_compensate}).  Intuitively, scaling the inputs to  Lev seems like a reasonable way to vary period because of the central role of levation in stepping, and it would be interesting to see if experimental manipulation of these signals produces matching effects.  We also found that uniform scaling of the external input strength parameters $\{ g_{ton,i} \}_i$
modulates period through changes in stance phase duration.  This result is reassuring, since sensory factors perceived elsewhere in the neural circuitry should be able to alter stepping rate through such a pathway. We also find it interesting that a uniform scaling can be effective, given the high degree of non-uniformity in the coupling strengths and interaction patterns across our model circuit.  It is true that in our model, the range of external input scaling factors that maintained (P) was relatively small; we expect that this limitation may reflect the fact that the parameter ranges we used were not carefully tuned to reflect specific biological measurements, but rather to achieve functional outputs through biologically-based mechanisms.

While we explored certain variations on our baseline model, such as the inclusion of motoneurons, of inter-joint inhibitory signaling in addition to inter-joint excitation, and an alternative parameter tuning to achieve transitions through release instead of escape, none of these yielded a significant improvement. It is possible that the use of a higher-dimensional motoneuron model could enhance the robustness of model outputs, at the cost of increasing the number of parameters to tune and analyze, and this represents a possible direction for future work.  Intra-segment inhibition could, in theory, contribute to the experimentally observed influences between joint component outputs \cite{bidaye2018}. Our results show that these synapses do not compromise the circuit's rhythmic outputs and could slightly enhance their robustness to parameter variations, but we feel that the more important take-away here is that such inhibitory interactions are not necessary for rhythm production and tuning.  When we switched from escape transitions to release transitions (Section \ref{sec:release}), in contrast, we found that inter-segment inhibition became necessary for successful rhythmicity consistent with (P). In fact, this finding is natural, because changes in inhibition are more impactful than changes in excitation on the existence and timing of the transitions from the active to the silent phase that underlie release.  Nonetheless, we did not find parameters that support a truly stable rhythm with release transitions, and hence our model predicts that escape is the transition mechanism underlying stick insect locomotor rhythm generation.  With transitions by escape, a general increase in excitability of the circuit will lead to a smaller period \cite{skinner1994moa,curtu,rubin2009}, and this property provides a natural way to test this prediction.  On other possibility worth mentioning, however, is that the metastable release rhythm could, in fact, be stable enough for real-world conditions, where changes in inputs and in external factors could induce ongoing circuit adjustments.  Testing the persistence of the rhythm under controlled conditions would presumably clarify the relevance of this scenario.

Obviously, our model is a highly simplified abstraction of the actual neural circuit underlying stick insect locomotion, the components of which are not even fully known at this point.  Simplification has allowed us to focus on the roles of specific model features and to derive predictions about possible rhythm generation and control mechanisms that would not have been possible with a more complicated model.  On the modeling side, one simplification we have made is that we have focused exclusively on non-spiking interneurons (NSIs).  While there are likely spiking neurons that also contribute to the dynamics we consider, experiments have established that NSIs contribute to locomotor rhythm generation and that their outputs represent a reliable indicator of the rhythm and project to associated motoneurons \cite{ludwar2005,vonuckermann2009}.  For the NSIs themselves, other voltage-gated currents that we have ignored could contribute to the dynamics; in this vein, it is important to keep in mind that models that achieve similar escape transitions through other electrophysiological mechanisms will yield similar results regarding effects of parameter modulations.   We have also ignored a number of factors from outside of the rhythm generation circuit that could influence its dynamics and outputs, including feedback signals not represented within the basic neuronal influence diagram as well as coupling between limbs and other biomechanical factors (e.g., \cite{borgmann2009,berendes2016,bidaye2018}), which have been included in some past models that were less focused on mechanistic analysis \cite{ekeberg2004,dauntoth,Amizare_Srivastava_Holmes_2018, Amizare_Holmes_2019, toth2012,toth2019,Proctor10,KukProc09} and should certainly be considered in future studies.

Beyond the model construction, we have focused on a specific output pattern, which is idealized and does not represent the variability found in real locomotor rhythms.  A possible direction of future research would be to analyze how small variations in parameters due to neuromodulation or other effects could account for experimentally observed output variability. This issue naturally connects to the broader question of how stick insects achieve other stepping patterns including those associated with maneuvers such as turns and backwards stepping, which remains for future modeling work.  Our model exhibits robustness to some parameter variations but not to others; similarly, it allows for effective tuning of stepping period through certain parameter modulations that we have discussed, whereas the rhythm period remains invariant over the ranges of other parameters where the rhythm persists.  These results represent predictions for testing with experiments and with simulations of more complex models.  The invariance of the period could be a flaw that goes away with the inclusion of more flexible dynamic mechanisms, or it could be a feature that allows useful behavior to persist in the face of metabolic, environmental, and other fluctuations.  

Overall, our work shows that a highly reduced model neuron circuit can produce an idealized, persistent stick insect locomotor rhythm as long as its parameters are tuned (a) to yield phase transitions through escape mechanisms, fueled by excitatory signaling, and (b) to ensure that activations of specific neurons occur specifically through either FTM, EE, or ghost mechanisms, or combinations thereof, to achieve appropriate transition timing.  These findings should bring attention to the relation between transition delays in neuronal rhythms and the  dynamic mechanisms that underlie them, which, as we show, have significant implications for robustness of model output pattern and modulation of output period with variation of specific circuit parameters.

\section*{Acknowledgment}
The authors acknowledge support from NSF award 2037828 and The Simons Foundation MPS--TSM--00008005 (ZA) and from NSF award DMS1951095 (JR).  We thank Silvia Daun for useful discussions in launching this project.

%~~~~~~~~~~~~~~~~~~~~~~~~~~~~~~~~
%\bibliographystyle{IEEEtran}
%\bibliography{nsf2024}


%~~~~~~~~~~~~~~~~~~~~~~~~~~~~~~~~~~
\section{Appendix}\label{Sec:appendix}
\appendix 

\subsection*{Parameter values used in simulations}

In the following tables, we organize all the parameters used in the models and simulations throughout the paper.
\begin{table}[h!]
\centering
\caption{Model parameters for interneuron-motoneuron units described in Equations~\eqref{eq:single_CPG_MN}-\eqref{eq:tau_h}}
\label{table:params}
\begin{tabular}{lccc}
\hline
Parameter & Equation & Escape & Release \\
\hline\hline
$C_m$ & \eqref{eq:single_v} and \eqref{eq:single_w}& 0.21 & 0.21 \\
 \hline
$g_{NaP}$  &\eqref{eq:I_Nap_IN}-\eqref{eq:I_Nap_MN}& 10.0 & 10.0 \\
$E_{NaP}$&\eqref{eq:I_Nap_IN}-\eqref{eq:I_Nap_MN}& 50.0 & 50.0 \\\hline
$g_{Leak}$ &\eqref{eq:I_leak} &2.8 &4.0 \\
$E_{Leak}$ &\eqref{eq:I_leak} & -65.0 & -60.0  \\
\hline
$\theta_m$ &\eqref{eq:m_infty}& -37 & -32.0 \\
$\sigma_m$ &\eqref{eq:m_infty}& -6 & -5.5 \\
\hline
$\thethin$ &\eqref{eq:h_infty_IN}&  -30 & -50 \\
$\sighin$ & \eqref{eq:h_infty_IN}& 7 & 12\\
\hline
$\thethmn$ & \eqref{eq:h_infty_MN}&{-7 \& -18} & -50 \\
$\sighmn$  &\eqref{eq:h_infty_MN}& 7 & 12\\
\hline
$\theta_{\tau}$ &\eqref{eq:tau_h} &-30 &-30 \\
$\sigma_{\tau}$   &\eqref{eq:tau_h}& 7& 7 \\
$\epsilon$  & \eqref{eq:tau_h}& 0.095& 0.095\\  \hline
$g_{ton}$    &\eqref{eq:I_app_in_mn} & \text{varies in Table~\ref{table:params_intrinsic}} & \text{varies in Table~\ref{table:params_intrinsic}} \\
$\Eexc$   & \eqref{eq:I_app_in_mn}&  0.0 & 10\\
\hline
\end{tabular}
\end{table}

\begin{table}[h!]
\centering
\caption{Excitatory and inhibitory synaptic parameters given in Equations~\eqref{eq:I_syn}-\eqref{eq:s_infty3}}
\label{table:params_inh_exc_coupling}
\begin{tabular}{lccc}
\hline
Parameter & Equation & Escape & Release \\
\hline\hline
$\gsynin$  &\eqref{eq:I_syn} & varies in Table~\ref{table:params_syn_inh_coupling} & varies in Table~\ref{table:params_syn_inh_coupling}\\
$\gsynmn$   &\eqref{eq:I_syn_MN} & 10 &10\\
$\Einh$    &\eqref{eq:I_syn} & -80 & -80\\
$\theta_{inh}$   &\eqref{eq:s_infty1} & -43& -30\\
$\sigma_{inh}$  &\eqref{eq:s_infty1} &-0.1&-0.1 \\ \hline
$\gexji$  &\eqref{eq:I_syn_exc} & varies in Table~\ref{table:params_exc_coupling} & varies in Table~\ref{table:params_exc_coupling}\\
$\Eexc$   &  \eqref{eq:I_syn_exc}&  0.0 & 10\\
$\theta_{exc}$   &\eqref{eq:s_infty2} &-37 & -37 \\
$\sigma_{exc}$   &\eqref{eq:s_infty2} &-6 &-6 \\ 
\hline
$\ginji$  &\eqref{eq:I_syn_inh} & varies in Table~\ref{table:params_inh_coupling} & varies in Table~\ref{table:params_inh_coupling}\\
$\Einh$    & \eqref{eq:I_syn_inh} & -80 & -80\\
$\hat\theta_{inh}$   &\eqref{eq:s_infty3} & -25& -25\\
$\hat\sigma_{inh}$   &\eqref{eq:s_infty3} &-0.1& -0.1\\\hline
\end{tabular}
\end{table}

\begin{table}[h!]
\centering
\caption{Constant conductance parameters of tonic currents applied to each IN as defined in  \eqref{eq:I_app_in_mn}}
\label{table:params_intrinsic}
\begin{tabular}{lcccc}
\hline
Parameter &  only IN, Escape &  IN-MN, Escape  & only IN, Release\\
\hline\hline
$g_{ton}$ & Baseline 1 & Baseline 2& Baseline 3\\
$\gappp$ &  0.17 &  0.1906  &0.25\\
$\gappr$  &   0.22 &0.2046 &0.75\\
$\gappl$ &  0.18 & 0.1772 &0.3\\
$\gappd$ &  0.19&  0.1767 &0.7\\
$\gappe$ &  0.19 &  0.2046 &1.0\\
$\gappf$ &  0.19 & 0.1772 &0.3\\
\hline
\end{tabular}
\end{table}

\begin{table}[h!]
\centering
\caption{Inhibitory coupling strengths of reciprocally inhibitory units in each joint as defined in  \eqref{eq:I_syn_IN}}
\label{table:params_syn_inh_coupling}
\begin{tabular}{lcccc}
\hline
Parameter &  only IN, Escape &  IN-MN, Escape  & only IN, Release\\
\hline\hline\medskip
$\gsynin$ &   Baseline 1 & Baseline 2& Baseline 3\\\medskip
$\gsynRP$  & 1 &1.0914 & 1    \\\medskip
$\gsynPR$  &   1.02 &1.0914& 1 \\\medskip
$\gsynDL$ &   1 &1.0404& 1 \\\medskip
$\gsynLD$ &  1&1.0404& 1  \\\medskip
$\gsynFE$ &  1.03 &1.0506& 1  \\\medskip
$\gsynEF$ &    1.03 & 1.0506& 1\\
\hline
\end{tabular}
\end{table}

\begin{table}[ht]
\centering
\caption{The strengths of excitatory synaptic coupling across the joints, as described in \eqref{eq:I_syn_exc}}
\label{table:params_exc_coupling}
\begin{tabular}{lcccc}
\hline
Parameter &  only IN, Escape &  IN-MN, Escape  & only IN, Release\\
\hline\hline\medskip
$\gexji$ &  Baseline 1 & Baseline 2 & Baseline 3\\\medskip
$\gexLP$ &  0.019 & 0.0346 & 0.1\\\medskip
$\gexLE$ &  0.1   & 0.2770 & 0.3\\\medskip
$\gexED$ &  0.025 & 0.0554 & 0.4\\\medskip
$\gexDR$ &  0.01  & 0.1385 & 0.1\\\medskip
$\gexDF$ &  0.016 & 0.0582 & 0.005\\\medskip
$\gexFF$ &  0.04  & 0.0554 & 0.1  \\\medskip
$\gexFL$ &  0.01  & 0.0443 &  0.1\\\medskip
$\gexFE$ &  0.008 & 0.0305 & 0.1 \\
\hline
\end{tabular}
\end{table}

\begin{table}[ht]
\centering
\caption{The strengths of inhibitory synaptic coupling across the joints, as described in \eqref{eq:I_syn_inh}}
\label{table:params_inh_coupling}
\begin{tabular}{lcccc}
\hline
Parameter &   only IN, Escape &  IN-MN, Escape  & only IN, Release\\
\hline\hline\medskip
$\ginji$ &   Baseline 1 & Baseline 2 & Baseline 3\\\medskip
$\ginLR$ &  0.019 & 0.0 & 0.1\\\medskip
$\ginLF$ &  0.1   & 0.0 & 0.1\\\medskip
$\ginEL$ &  0.025 & 0.0 &0.075\\\medskip
$\ginDP$ &  0.01  & 0.0 & 0.19 \\\medskip
$\ginDE$ &  0.016 & 0.0 & 0.013\\\medskip
$\ginFE$ &  {0.04}  & 0.0 & 0.0 \\\medskip
$\ginFD$ & 0.01  & 0.0 & 0.1 \\\medskip
$\ginFF$ &  0.008  & 0.0 & 0.1 \\
\hline
\end{tabular}
\end{table}

\end{document}